\documentclass[useAMS,usenatbib]{mn2e}
\usepackage{epsfig}
\usepackage{graphicx}
\usepackage{journal_names}
\usepackage[english]{babel}
\usepackage{fancyhdr,txfonts} 
\usepackage{hyperref}

\usepackage{natbib}
\usepackage{graphicx}
\usepackage{sidecap}

\usepackage{amssymb}
\usepackage{color}
\usepackage{wasysym}
\def\gsim{\;\rlap{\lower 2.5pt\hbox{$\sim$}}\raise 1.5pt\hbox{$>$}\;}
\def\lsim{\;\rlap{\lower 2.5pt\hbox{$\sim$}}\raise 1.5pt\hbox{$<$}\;}

\title[The IGM thermal history down to z =1.5]{The thermal history of the intergalactic medium down to redshift z=1.5: a new curvature measurement}
\author[Boera et al.]{\noindent
Elisa Boera$^{1}$\thanks{E-mail: eboera@swin.edu.au}, Michael T.~Murphy$^{1}$, George D.~Becker$^{2}$, James S.~Bolton$^{3}$
\\~\\
$^1$ Centre for Astrophysics and Supercomputing,  Swinburne University of Technology, Hawthorn, Victoria 3122, Australia\\
$^2$ Kavli Institute for Cosmology and Institute of Astronomy, Madingley Road, Cambridge, CB3 0HA\\
$^3$ School of Physics and Astronomy, University of Nottingham, University Park, Nottingham, NG7 2RD\\
}

\begin{document}

\date{Accepted 2014 March 30.  Received 2014 March 29; in original form 2014 February 16 }

\pagerange{\pageref{firstpage}--\pageref{lastpage}} \pubyear{2014}

\maketitle

\label{firstpage}

\begin{abstract}
According to the photo-heating model of the intergalactic medium (IGM), He II reionization is expected to affect its thermal evolution. Evidence for additional energy injection into the IGM has been found at $3\lesssim z\lesssim4$, though the evidence for the subsequent fall-off below $z\sim2.8$ is weaker and depends on the slope of the temperature--density relation, $\gamma$. Here we present, for the first time, an extension of the IGM temperature measurements down to the atmospheric cut-off of the H I Lyman-$\alpha$ forest at $z\simeq1.5$. Applying the curvature method on a sample of 60 UVES spectra we investigated the thermal history of the IGM at $z<3$ with precision comparable to the higher redshift results. We find that the temperature of the cosmic gas traced by the Ly-$\alpha$ forest [$T(\bar{\Delta})]$ increases for increasing overdensity from $T(\bar{\Delta})\sim 22670$ K to 33740 K in the redshift range $z\sim2.8-1.6$. Under the assumption of two reasonable values for $\gamma$, the temperature at the mean density ($T_{0}$) shows a tendency to flatten at $z\lesssim 2.8$. In the case of $\gamma\sim1.5$, our results are consistent with previous ones which indicate a falling $T_{0}$ for redshifts $z\lesssim2.8$. Finally, our $T(\bar{\Delta})$ values show reasonable agreement with moderate blazar heating models.

\end{abstract}


\section{Introduction}
Starting from a very hot plasma made of electrons and protons after the Big Bang, to the gas that now fills the space between galaxies, the Intergalactic medium (IGM) has been one of the main ``recorders'' of the different phases of evolution of the Universe. Changes in the thermodynamic state and chemical composition of this gas reflect the conditions for the formation and the evolution of the structures that we can observe today. 
 In particular, the IGM thermal history can be an important source of information about reionizing processes that injected vast amounts of energy into this gas on relatively short timescales. For this reason considerable efforts have been made to find any ``footprint" of either H I ($z <6$) or He II ($z<3$) reionization. Because the ionisation potential of He II (from He II to He III) is 54.4 eV and fully ionized helium recombines more than 5 times faster than ionized hydrogen, this second reionization event should have begun later, after the reionization of hydrogen and He I ($11\gtrsim z \gtrsim 6$, 
\citealt{LL11};
 \citealt{Fan06}; 
\citealt{Becker01}) when quasars started to dominate the UV background (\citealt{MiraldaescudeQ00}). Theoretically, the much harder photons from quasars would have been able to fully ionize He II around redshifts $3\lesssim z  \lesssim 4.5$ but these estimates change depending on assumptions about the abundance of quasars (QSOs) and the hardness of their spectra (\citealt{Meiksin05}). While the direct observation, through the detection of the ``Gunn--Peterson effect'', recently suggests the end of the He II reionization at $z\sim 2.7$ (e.g. \citealt{Shull10};\citealt{Worseck11}; \citealt{Syphers13}; \citealt{Syphers14}; ), any current constraint on the physics of this phenomenon is limited by the cosmic variance among the small sample of ``clean''   lines of sight, those along which the He II Lyman-$\alpha$ transition is not blocked by higher-redshift HI Lyman limit absorption. For this reason indirect methods have been developed to obtain a detailed characterization of the He II reionization.

It is predicted that the IGM should be reheated by photo-ionization heating during He II reionization and, because its cooling time is long, the low density gas retains some useful memory  of when and how it was reionized. In fact, at the mean density of the IGM the characteristic signature of reionization is expected to be a peak: a gradual heating followed by cooling due to adiabatic expansion (e.g. \citealt{McQuinn09}).
In the last decade, the search for this feature and the study of the thermal history of the IGM as a function of redshift have been the objectives of different efforts, not only to verify the theoretical prediction and constrain the timing of He II reionization, but also to obtain information on the nature of the ionizing sources and on the physics of the related ionizing mechanisms. To obtain measurements of the temperature of the IGM, studying the absorption features of the HI Lyman-$\alpha$ forest has proven to be a useful method so far. The widths of these lines are sensitive to thermal broadening but are also affected by Hubble broadening and peculiar velocities. Cosmological simulations are therefore required to characterize the large scale structure and bulk motion of the IGM (\citealt{Meiksin10}), before the gas temperature can be determined. 

Previous efforts to extract information on the thermal state of the cosmic gas from the Lyman-$\alpha$ forest can be divided into two main approches: the study of individual absorption features and the quantification of the absorption structures with a global statistical analysis of the entire forest.
The first method consists of decomposing the spectra into a set of Voigt profiles. \citet{Schaye00} found evidence using this technique for an increase in the IGM temperature consistent with He II reionization at $z\simeq3$. In contrast, \citet{McDonald01} found a constant temperature over $z\sim2-4$. A characterization of the flux probability distribution (PDF) based on pixel statistics has also been used to analyse the forest and extract information from the comparison with theoretical models (\citealt{Bolton08}; \citealt{Calura12}). However, the PDF is sensitive to a range of systematic effects, including the placement of the unabsorbed continuum. A further approach is to use wavelet analysis to characterize the Ly$\alpha$ line-widths distribution in terms of discrete wavelets. \citet{Theuns02} and \citet{Lidz10} found evidence using this technique for He II reionization completing near $z\sim3.4$, but with large statistical uncertainties. In the recent work of \citet{Garzilli12}, the PDF and the wavelet decomposition methods have been compared and tested on Lyman-$\alpha$ spectra at low redshift. While the results are in formal agreement with previous measurements, the uncertainties are still large and there is a mild tension between the two analyses.

Recently, \citet{Becker11} developed a statistical approach based on the flux curvature. This work constrained the temperature over $2\lesssim z\lesssim 4.8$ of a ``optimal" or ``characteristic"  overdensity, which evolves with redshift. The error bars were considerably reduced compared to previous studies, partially at the expense of determining the temperature at a single density only, rather than attempting  to constrain  the temperature--density relation. Some evidence  was found for a gradual reheating  of the IGM over $3 \lesssim z \lesssim 4$ but with no clear evidence for a temperature peak.  Given these uncertainties, the mark of the He II reionization still needs a clear confirmation. Nevertheless, the curvature method is promising because it is relatively robust to continuum placement errors: the curvature of the flux is sensitive to the shape of the absorption lines and not strongly dependent on the flux normalization. Furthermore, because it incorporates the temperature information from the entire Lyman-$\alpha$ forest, this statistic has the advantage of using more of the available information, as opposed to the line-fitting method which relies on selecting lines that are dominated by thermal broadening. 

An injection of substantial amounts of thermal energy is predicted to result in both an increase in the IGM temperature and a change in the temperature--density (T--$\rho$) relation.  The detailed study of this process has to take into consideration the effects of the IGM inhomogeneities driven by the diffusion and percolation of the ionized bubbles around single sources, and currently constitutes an important object of investigation through hydrodynamical simulations (e.g. \citealt{Compostella13}).  In the simplest scenario, for gas at overdensities $\Delta\lesssim 10$ ($\Delta=\rho / \bar{\rho}$, where $\bar{\rho}$ is the mean density of the IGM), the temperature is related to the density with a power-law relation of the form: 
\begin{equation} 
T(\Delta)=T_{0}\Delta^{\gamma -1} ,
\end{equation}
 where $T_{0}$ is the temperature at the mean density (\citealt{HuiGnedin1997}; \citealt{Valageas02}). The evolution of the parameters $T_{0}$ and $\gamma$ as a function of redshift then describes the thermal history of the IGM. A  balance between photo-heating and cooling due to adiabatic expansion of the Universe will asymptotically produce a power law with $\gamma=1.6$ (\citealt{HuiGnedin1997}). During the reionization the slope is expected to flatten temporarily before evolving back to the asymptotic value. Possible evidence for this flattening at $z\simeq3$, seems to be consistent with He II reionization occurring around this time (e.g., \citealt{Ricotti00}; \citealt{Schaye00}). 

Some analyses of the flux PDF have indicated that the T--$\rho$ relation may even become inverted (e.g., \citealt{Becker07}; \citealt{Bolton08}; \citealt{Viel09}; \citealt{Calura12}; \citealt{Garzilli12}). However, the observational uncertainties in this measurement are considerable (see discussion in \citealt{Bolton13}). A possible explanation was suggested by considering radiative transfer effects (\citealt{Bolton08}). 
Although it appears difficult to produce this result considering only  He II photo-heating by quasars (\citealt{McQuinn09};\citealt{Bolton09}), a new idea of volumetric heating from blazar TeV emission predicts an inverted temperature--density relation at low redshift and at low densities. According to these models,  heating by blazar $\gamma$-ray emission would start to dominate at $z\simeq3$, obscuring the ``footprint" of He II reionization (\citealt{Chang12}; \citealt{Puchwein12}). Even if in the most recent analysis, with the line-fitting method (\citealt{Rudie13}; \citealt{Bolton13}), the inversion in the temperature--density relation has not been confirmed, a general lack of knowledge about the behaviour of the T--$\rho$ relation at low redshift ($z<3$) still emerges, accompanied with no clear evidence for the He II reionization peak.  A further investigation of the temperature evolution in this redshifts regime therefore assumes some importance in order to find constraints for the physics of the He II reionization and the temperature--density relation of the IGM.

The purpose of this work is to apply the curvature method to obtain new, robust temperature measurements at redshift $z<3$, extending the previous results, for the first time, down to the optical limit for the Lyman-$\alpha $ forest at $z\simeq1.5$. By pushing the measurement down to such a low redshift, we attempt to better constrain the thermal history in this regime, comparing the results with the theoretical predictions for the different heating processes. We infer temperature measurements by computing the curvature on a new set of quasar spectra at high resolution obtained from the archive of the UVES spectrograph on the VLT. Synthetic spectra, obtained from hydrodynamical simulations used in the analysis of \citet{Becker11} and extended down to the new redshift regime are used for the comparison with the observational data. Similar to \citet{Becker11}, we  constrain the temperature of the IGM at a characteristic overdensity, $\bar{\Delta}$, traced by the Lyman-$\alpha$ forest, which evolves with redshift. We do not attempt to constrain the T--$\rho$ relation, but we use fiducial values of the parameter $\gamma $ in Eq. 1 to present results for the temperature at the mean density, $T_{0}$.  

This paper is organised as follows. In Section 2 we present the observational data sample obtained from the VLT archive, while the simulations used to interpret the measurements are introduced in Section 3. In Section 4 the curvature method and our analysis procedure are summarized. In Section 5 we present the data analysis and we discuss the strategies applied to reduce the systematic uncertainties. The calibration and the analysis of the simulations is described in Section 6. The results are presented in Section 7 for the temperature at the characteristic overdensities and the temperature at  the mean density for different values of  $\gamma$. We discuss the comparison with theoretical models in Section 8, and conclude in Section 9. 

\section{THE OBSERVATIONAL DATA}\label{sec:obs}

In this work we used a sample of 60 quasar spectra uniformly selected on the basis of redshift, wavelength coverage and S/N in order to obtain robust results in the UV and optical parts (3100--4870\,\AA) of the spectrum where the Lyman-$\alpha$ transition falls for redshifts $1.5<z<3$. The quasars and their basic properties are listed in Table \ref{table:datatable}.  The spectra were retrieved from the archive of the UVES spectrograph on the VLT. In general, most spectra were observed with a slit-width $\lesssim 1\farcs0$ wide and on-chip binning of 2$\times$2, which provides a resolving power $R\simeq 50000$ (FWHM $\simeq$ 7 km/s); this is more than enough to resolve typical Lyman-$\alpha$ forest lines, which generally have FWHM $\gtrsim$ 15 km/s. The archival quasar exposures were reduced using the ESO UVES Common Pipeline Language software. This suite of standard routines was used to optimally extract and wavelength-calibrate individual echelle orders. The custom-written program {\sc uves\_popler}\footnote{{\sc uves\_popler} was written and is maintained by M.~T.~Murphy and is available at \url{http://astronomy.swin.edu.au/~mmurphy/UVES_popler}.}
was then used to combine the many exposures of each quasar into a single normalized spectrum on a vacuum-heliocentric wavelength scale. For most quasars, the orders were redispersed onto a common wavelength scale with a dispersion of 2.5 km/s per pixel; for 4 bright (and high S/N), $z\lesssim2$ quasars the dispersion was set to 1.5 km/s per pixel. The orders were then scaled to optimally match each other and then co-added with inverse-variance weighting using a sigma-clipping algorithm to reject `cosmic rays' and other spectral artefacts.

To ensure a minimum threshold of spectral quality and a reproducible sample definition, we imposed a ``S/N" lower limit of 24 per pixel for selecting which QSOs and which spectral sections we used to derive the IGM temperature.  A high S/N is, in fact, extremely important for the curvature statistic which is sensitive to the variation of the shapes of the Lyman-$\alpha$ lines: in low S/N spectra this statistic will be dominated by the noise and, furthermore, by narrow metal lines that are difficult to identify and mask.

The ``S/N'' cut-off of 24 per pixel was determined by using the hydrodynamical Lyman-$\alpha$ forest simulations discussed in Section \ref{sec:sims}. By adding varying amounts of Gaussian noise to the simulated forest spectra and performing a preliminary curvature analysis like that described in Sections \ref{sec:curvature} \& \ref{sec:analysis}, the typical uncertainty on the IGM temperature could be determined, plus the extent of any systematic biases caused by low S/N. It was found that a competitive statistical uncertainty of $\simeq 10$\% in the temperature could be achieved with the cut-off in ``S/N'' set to 24 per pixel, and that this was well above the level at which systematic biases become significant. However, in order for us to make the most direct comparison with these simulations, we have to carefully define ``S/N". In fact for the Lyman-$\alpha$ forest the S/N fluctuates strongly and so it is not very well defined. Therefore, the continuum-to-noise ratio, C/N, is the best means of comparison with the simulations. To measure this from each spectrum, we had first to establish a reasonable continuum.

\begin{figure}
\centering
\begin{tabular}{cc}
\includegraphics[width=0.4\textwidth]{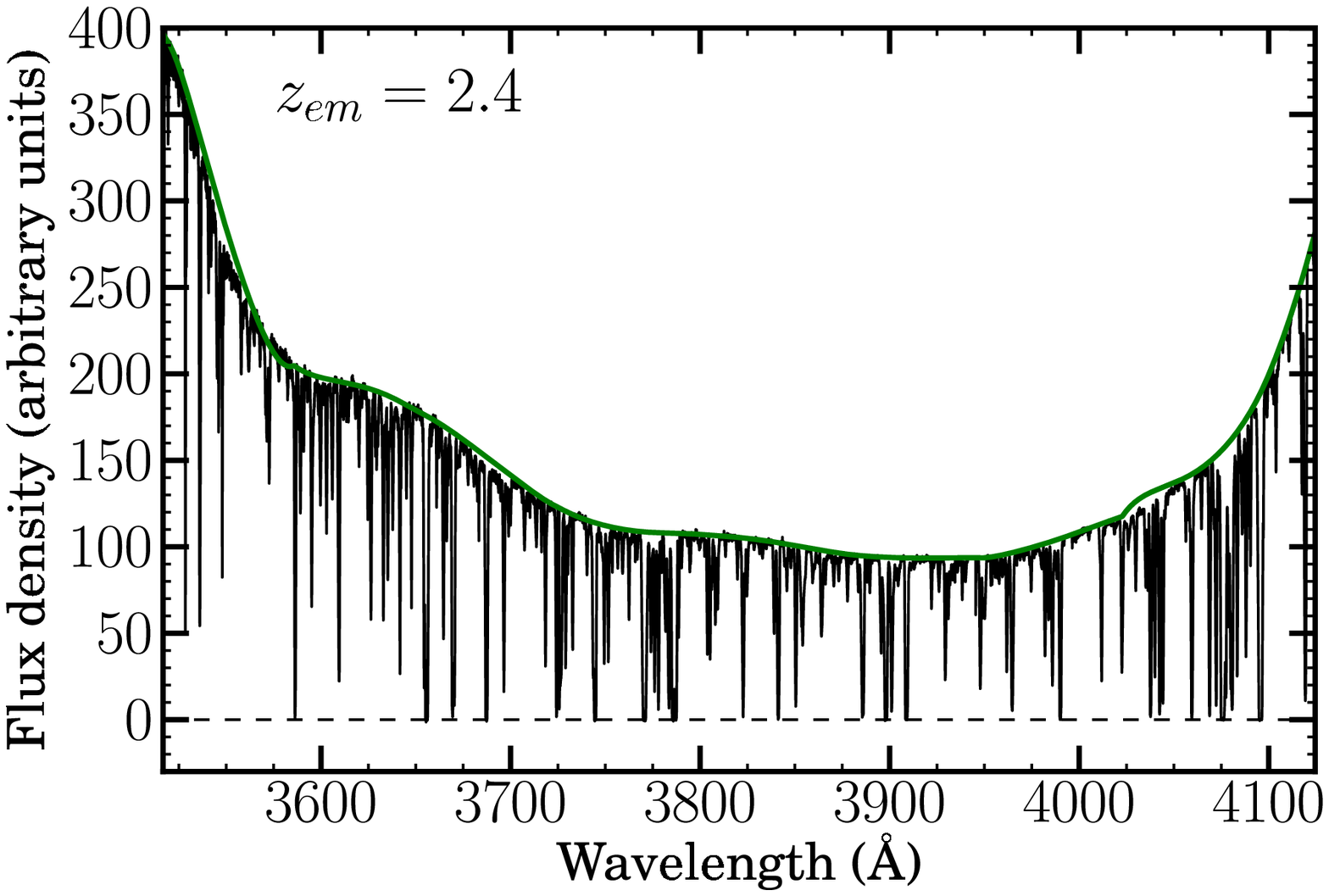}\\
\includegraphics[width=0.4\textwidth]{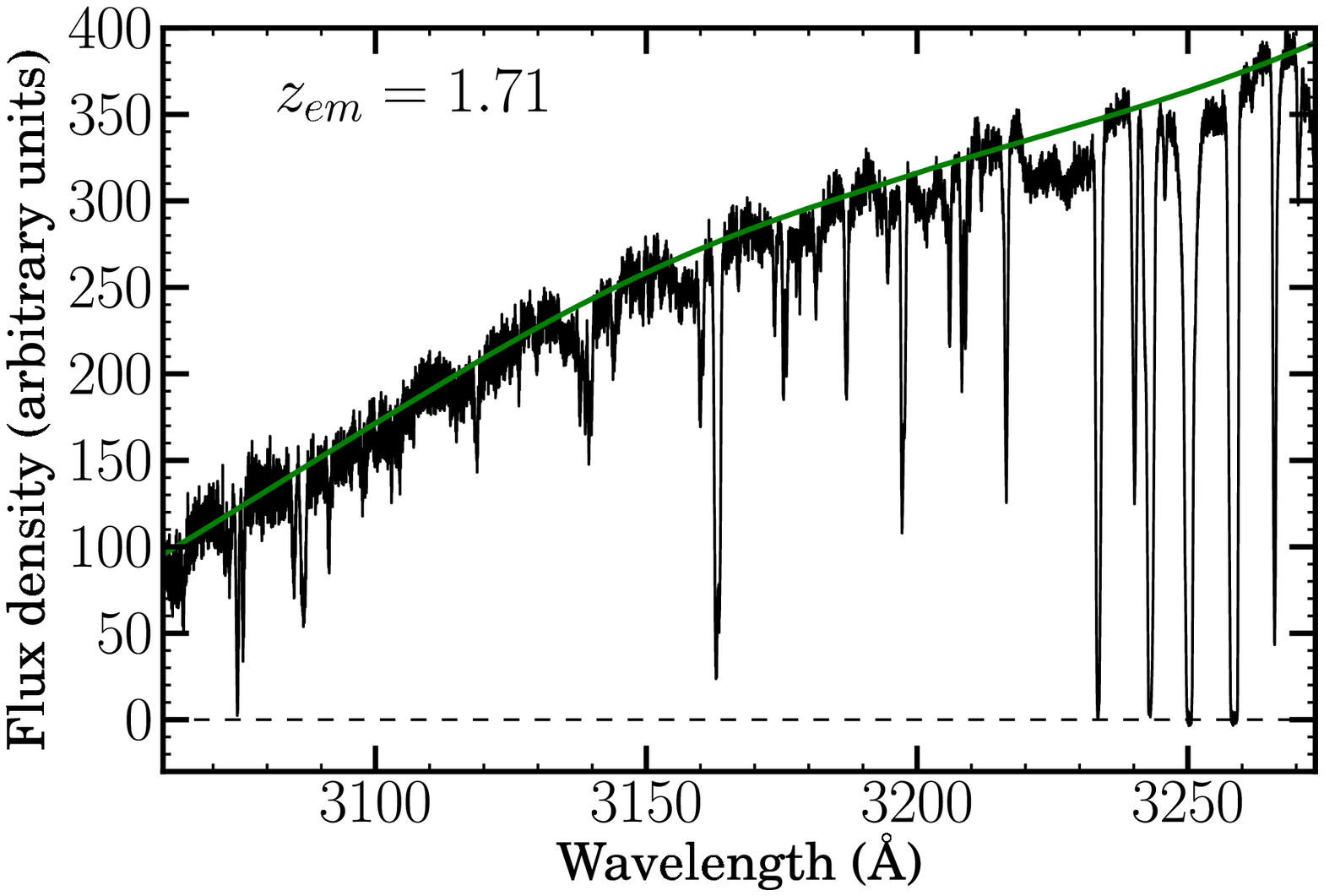}\\
\end{tabular}
\caption{\small Examples of continuum fit (green solid line) in Lyman-$\alpha$ regions for the quasar  J112442-170517 with $z_{em}= 2.40$ (Top panel) and  J051707-441055 at $z_{em}= 1.71$ (Bottom panel). The continuum fitting procedure  used is described in the text.}
\label{fig:forestex}
\end{figure}

The continuum fitting is a crucial aspect in the quasar spectral analysis and for this reason we applied to all the data a standard procedure in order to avoid systematic uncertainties due to the continuum choice. We used the continuum-fitting routines of {\sc uves\_popler} to determine the final continuum for all our quasar spectra. Initially, we iteratively fitted a 5th-order Chebyshev polynomial to overlapping 10000-km/s sections of spectra between the Lyman-$\alpha$ and Lyman-$\beta$ emission lines of the quasar. The initial fit in each section began by rejecting the lowest 50\,\% of pixels. In subsequent iterations, pixels with fluxes $\ge$3\,$\sigma$ above and $\ge$1\,$\sigma$ below the fit were excluded from the next iteration. The iterations continued until the `surviving' pixels remained the same in two consecutive iterations. The overlap between neighbouring sections was 50\,\% and, after all iterations were complete, the final continuum was formed by combining the individual continua of neighbouring sections with a weighting which diminished linearly from unity at their centres to zero at their edges. After this initial treatment of all spectra we applied further small changes to the fitting parameters after visually inspecting the results. In most cases, we reduced the spectral section size, the threshold for rejecting pixels below the fit at each iteration, and the percentage of pixels rejected at the first iteration to values as low as 6000 km/s, 0.8 $\sigma$ and 40\,\% respectively. In Figure \ref{fig:forestex} we show examples of continuum fits for Ly-$\alpha$ forest regions at different redshifts obtained with this method. This approach allowed us to avoid cases where the fitted continuum obviously dipped inappropriately below the real continuum, but still defined our sample with specific sets of continuum parameters without any further manual intervention, allowing a reproducible selection of the appropriate sample for this analysis. Furthermore, as described in Section 5, to avoid any systematics due to the large-scale continuum placement, we re-normalized each section of spectra that contributed to our results.

\begin{figure} \label{fig:histo}
\centering
\includegraphics[width=0.5\textwidth]{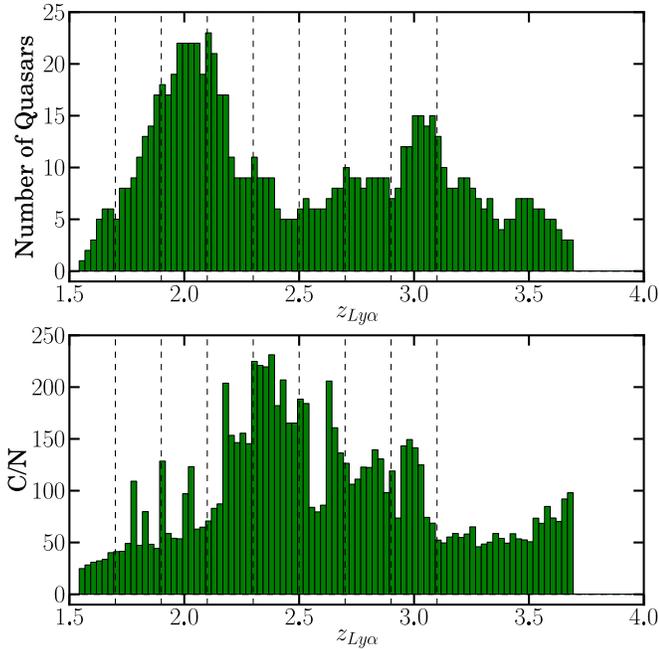} 
\caption{\small  Histograms showing our sample of quasar spectra with C/N$> 24$ in the Lyman- $\alpha$ forest region. Top panel: redshift distribution referred to the Lyman$-\alpha$ forest. Bottom panel: distribution of the continuum to noise ratio (C/N) in the forest region. The redshift bins of the histograms have been chosen for convenience of $\delta z=0.025$. The vertical lines divide the histograms in the redshift bins width of $\Delta z= 0.2$ in which the measurements (at $z\lesssim 3.1$) will be collected in the following analysis.}
\end{figure}

The redshift distribution of the Lyman-$\alpha$ forest ($z_{Ly\alpha}$) of the quasars in our selected sample is shown in Figure \ref{fig:histo}, where it is also reported their distribution of C/N in the same region. Due to instrumental sensitivity limits in the observation of bluest part of the optical Ly-$\alpha$ forest, it is more difficult to collect data with high C/N for $z_{Ly\alpha} < 1.7$. The general lower quality of these data and the lower number of quasars contributing to this redshift region  will be reflected in the results, causing larger uncertainties.

\begin{table} 
\caption{\small List of the QSOs used for this analysis. For each object we report the name (column 1) based on the J2000 coordinates of the QSO and the emission redshift (column 2). The redshift intervals associated with the Lyman-$\alpha$ absorption are also reported with the corresponding C/N level per pixel (columns 3, 4 \& 5). The sections of Lyman-$\alpha$ forest obtained from this sample were required to have a minimum C/N of 24 per pixel. The note below the table provides the ESO Program IDs that contributed to the spectra.} 
\label{table:datatable}
\centering \begin{tabular}{c c c c c c} 
\hline\hline
QSO\ & $z_{em}$\ & \multicolumn{2}{c}{$z_{Ly\alpha}$}  & C/N  \\

\multicolumn{1}{c}{} &
\multicolumn{1}{c}{} &
\multicolumn{1}{c}{$z_{start}$} &
\multicolumn{1}{c}{$z_{end}$} &
\multicolumn{1}{c}{} \\

\hline J092913$-$021446 & 1.6824 & 1.263 & 1.6824 &5$-$29\\ 
J051707$-$441055 & 1.71 & 1.286 & 1.71 & 10$-$54\\ 
J014333$-$391700 & 1.807 & 1.368 & 1.807 &  13$-$69 \\ 
J222756$-$224302 & 1.891 & 1.439 & 1.891 &   15$-$73\\ 
J013857$-$225447 & 1.893 & 1.440 & 1.893 & 16$-$52\\
J013105$-$213446 & 1.9 & 1.446 & 1.9 & 2$-$35\\
J005824$+$004113 & 1.92 & 1.463 & 1.92 &  10$-$43\\
J043037$-$485523 & 1.94 &  1.480 & 1.94 &  2$-$35\\
J010821$+$062327 & 1.96 &  1.497 & 1.96  & 35$-$56\\
J115944$+$011206 & 2.0001 & 1.531 & 2.0001 &  5$-$30\\
J115940$-$003203 & 2.035 & 1.560 & 2.035 &  6$-$35\\
J144653$+$011356 & 2.206 & 1.705 & 2.206 &  25$-$60\\
J024221$+$004912 & 2.0682 & 1.588 & 2.0682 &  8$-$26\\
J225719$-$100104 & 2.0797 & 1.598 & 2.0797 & 13$-$41\\
J133335$+$164903 & 2.084 & 1.602 & 2.084 &  26$-$91\\
J001602$-$001225 & 2.0869 & 1.604 & 2.0869 & 28$-$81\\
J124924$-$023339 & 2.1169 & 1.629 & 2.1169 & 15$-$50\\
J031009$-$192207 & 2.122 & 1.634 & 2.122 &  14$-$51\\
J031006$-$192124 & 2.144 & 1.652 & 2.144 &  15$-$62\\
J110325$-$264515 & 2.14500 & 1.653 & 2.145 & 50$-$270\\
J223235$+$024755 & 2.147 & 1.655 & 2.147 & 14$-$36\\
J042707$-$130253 & 2.159 & 1.665 & 2.159 &  11$-$42\\
J122310$-$181642 & 2.16 & 1.666 & 2.16 & 9$-$26\\
J121140$+$103002 & 2.193 & 1.694 & 2.193 & 16$-$63\\
J012417$-$374423 & 2.2004 & 1.700 & 2.2004 & 34$-$110\\
J145102$-$232930 & 2.215 & 1.712 & 2.215 &  60$-$140\\
J221531$-$174408 & 2.217 & 1.714 & 2.217 &  18$-$64\\
J024008$-$230915 & 2.223 & 1.719 & 2.223 &  80$-$182\\
J103921$-$271916 & 2.23 & 1.725 & 2.23 &  25$-$80\\
J212329$-$005052 & 2.2623 & 1.752 & 2.2623 &  70$-$106\\
J000344$-$232355 & 2.28 & 1.767 & 2.28 & 46$-$125\\
J045313$-$130555 & 2.3 & 1.784 & 2.3 &  37$-$48\\
J104032$-$272749 & 2.32 & 1.801 & 2.32 & 32$-$43\\
J211927$-$353740 & 2.341  & 1.818 & 2.341 & 16$-$50\\
J112442$-$170517 & 2.4 & 1.868 & 2.4  & 125$-$305 \\
J115122$+$020426 & 2.401 & 1.869 & 2.401 &  17$-$55\\
J222006$-$280323&  2.406 & 1.873 & 2.406  & 80$-$180\\
J011143$-$350300 & 2.41 & 1.877 & 2.41 &  67$-$130\\
J033106$-$382404 & 2.423 & 1.888 & 2.423 &  41$-$72\\
J120044$-$185944 & 2.448 & 1.909 & 2.448 &  62$-$108\\
J234628$+$124858 & 2.515 & 1.965 & 2.515 &  40$-$42\\
J015327$-$431137 & 2.74 & 2.155 & 2.74 &  75$-$130\\
J235034$-$432559 & 2.885 & 2.277 & 2.885 & 140$-$248\\
J040718$-$441013 & 3.0 &  2.37 & 3.0  & 96$-$135\\
J094253$-$110426 & 3.054 & 2.420 & 3.054 &  63$-$129\\
J042214$-$384452 & 3.11 &  2.467 & 3.11 &   98$-$160\\
J103909$-$231326 & 3.13 & 2.484 & 3.13 &  59$-$80\\
J114436$+$095904 & 3.15 & 2.501 & 3.15 & 25$-$43\\ 
J212912$-$153841 & 3.268 & 2.601 & 3.268 & 105$-$190\\
J233446$-$090812 & 3.3169 & 2.642 & 3.3169 &  40$-$46\\
J010604$-$254651 & 3.365 & 2.682 & 3.365 &  27$-$43\\
J014214$+$002324 & 3.3714 & 2.688 & 3.3714 & 28$-$35\\
J115538$+$053050 & 3.4752 & 2.775 & 3.4752 & 43$-$64\\
J123055$-$113909 & 3.528 & 2.820 & 3.528 &  18$-$39\\
J124957$-$015928 & 3.6368 &  2.912 & 3.6368 &  52$-$87\\
J005758$-$264314 & 3.655 & 2.927 & 3.655 & 74$-$90\\

\hline 
\end{tabular}

\end{table}

\addtocounter{table}{-1}
\begin{table} 
\caption{\small $-$ Continued.} 
\centering \begin{tabular}{c c c c c c} 
\hline\hline
QSO\ & $z_{em}$\ & \multicolumn{2}{c}{$z_{Ly\alpha}$}  & C/N  \\

\multicolumn{1}{c}{} &
\multicolumn{1}{c}{} &
\multicolumn{1}{c}{$z_{start}$} &
\multicolumn{1}{c}{$z_{end}$} &
\multicolumn{1}{c}{} \\

\hline

J110855$+$120953 & 3.6716 & 2.941 & 3.671 &  33$-$36\\
J132029$-$052335 & 3.70 & 2.965 & 3.70 &  42$-$81\\
J162116$-$004250 & 3.7027 & 2.967 & 3.7027 &  109$-$138\\
J014049$-$083942 & 3.7129 & 2.976 & 3.7129 & 37$-$38\\

\hline 
\end{tabular}

{\footnotesize ESO Program IDs: 60.A-9022, 60.A-9207, 60.O-9025, 65.O-0063, 65.O-0158, 65.O-0296, 65.O-0299, 65.O-0474, 65.P-0038, 65.P-0183, 66.A-0133, 66.A-0212, 66.A-0221, 66.A-0624, 67.A-0022, 67.A-0078, 67.A-0146, 67.A-0280, 67.B-0398, 68.A-0170, 68.A-0216, 68.A-0230, 68.A-0361, 68.A-0461, 68.A-0570, 68.A-0600, 68.B-0115, 69.A-0204, 69.B-0108, 70.A-0017, 70.B-0258, 71.A-0066, 71.A-0067, 71.B-0106, 71.B-0136, 072.A-0100, 072.A-0346, 072.A-0446, 072.A-0446, 072.B-0218, 073.B-0420, 073.B-0787, 075.A-0464, 075.B-0190, 076.A-0376, 076.A-0463, 076.A-0860, 077.A-0646, 078.A-0003, 079.A-0108, 079.A-0251, 079.A-0303, 079.A-0404, 079.B-0469, 080.A-0014, 080.A-0482, 080.A-0795, 081.A-0242, 081.B-0285, 166.A-0106, 267.A-5714, 273.A-5020.}
 
 \end{table}


\section{ THE SIMULATIONS}\label{sec:sims}
To interpret our observational results and extract temperature constraints from the analysis of the Lyman-$\alpha$ forest, we used synthetic spectra, derived from hydrodynamical simulation and accurately calibrated to match the real data conditions.
We performed a set of hydrodynamical simulations that span a large range of thermal histories, based on the models of \citet{Becker11} and extended to lower redshifts ($z<1.8$)  to cover the redshift range of our quasar spectra. The simulations were obtained with the parallel smoothed particle hydrodynamics code GADGET-3 that is the updated version of {\sc gadget-2} (\citealt{Springrl05}) with initial conditions constructed using the transfer function of \citet{Eisenstein99} and adopting the cosmological parameters $\Omega_{m}=0.26$, $\Omega_{\Lambda}$=0.74, $\Omega_{b}h^{2} =0.023$, $h=0.72$, $\sigma_{8}=0.80$, $n_{s}=0.96$, according to the cosmic microwave background constraints of \citet{Reichardt09} and \citet{Jarosik11}.  
The helium fraction by mass of the IGM is assumed to be $Y=0.24$ (\citealt{Olive04}). Because the bulk of the Lyman-$\alpha$ absorption corresponds  to overdensities $\Delta=\rho / \bar{\rho} \lesssim 10$, our analysis will not be affected by the star formation prescription, established only for gas particles with overdensities $\Delta >10^{3}$ and temperature T$< 10^{5}$K.

Starting at $z=99$ the simulations describe the evolution of both dark matter and gas using $2 \times 512^{3}$ particles with a gas particle mass of $9.2\times 10^{4} M_{\odot}$ in a periodic box of 10 comoving $h^{-1}$ Mpc. Instantaneous hydrogen reionization is fixed at $z=9$. From the one set of initial conditions, many simulations are run, all with gas that is assumed to be in the optically thin limit and in ionisation equilibrium with a spatially uniform ultraviolet background from \citet{Haardt01}. However, the photoheating rates, and so the corresponding values of the parameters $T_{0}$ and $\gamma$ of Equation 1, were changed between simulations. In particular, the photo-heating rates from \citet{Haardt01}  ($\epsilon_{i}^{HM01}$) for the different species ($i$=[H I, He I, He II]) have been rescaled using the relation $\epsilon_{i}=\zeta\Delta^{\xi}\epsilon_{i}^{HM01}$ where $\epsilon_{i}$ are the adopted photo-heating rates and $\zeta$ and $\xi$ are constants that change depending on the thermal history assumed. Possible bimodality in the temperature distribution at fixed gas density, observed in the simulations of \citet{Compostella13} in the early phases of the He II reionization, has not been taken into consideration in our models. We assume, in fact, that the final stages of He II reionization at $z<3$, when the IGM is almost completely reionized, can be described in a good approximation by a single temperature--density relation and are not affected anymore by the geometry of the diffusion of ionized bubbles. 

Our models do not include galactic winds or possible outflows from AGN. However, these are expected to occupy only a small proportion of the volume probed by the synthetic spectra and so they are  unlikely to have an important effect on the properties of the Lyman-$\alpha$ forest (see e.g. \citealt{Bolton08} and also \citealt{Theuns02} for a discussion in the context of the PDF of the Lyman-$\alpha$ forest transmitted fraction where this has been tested).         
  
A summary of the simulations used in this work is reported in Table \ref{table:simulations}. We used different simulation snapshots that covered the redshift range of our quasar spectra ($1.5\lesssim z\lesssim 3$) and, to produce synthetic spectra of the Ly-$\alpha$ forest, 1024 randomly chosen ``lines of sight''  through the simulations were selected at each redshift. To match the observational data, we needed to calibrate the synthetic spectra with our instrumental resolution, with the same H I Ly$\alpha$ effective optical depth and the noise level obtained from the analysis of the real spectra (see Section 6.1).

\begin{table} 
\caption{\small Parameters corresponding to the different simulations used in this work. For each simulation we report the name of the model (column 1), the constants used to rescale the photo-heating rates for the different thermal histories (columns 2 \& 3), the temperature of the gas at the mean density at $z=3$ (column 4) and the power-law index of the T--$\rho$ relation at $z=3$ (column 4). } 
\centering \begin{tabular}{c c c c c } 
\hline\hline
Model\ &  $\zeta$\ & $\xi$\ & $T_{0}^{z=3}$[K]\ & $\gamma^{z=3}$\ \\
\hline 
A15 & 0.30 & 0.00 & 5100 & 1.52\\
B15 &  0.80 & 0.00 & 9600 & 1.54\\
C15 & 1.45 & 0.00 & 14000 & 1.54\\
D15 & 2.20 & 0.00 & 18200 & 1.55\\
E15 & 3.10 & 0.00 &  22500 & 1.55\\
F15 & 4.20 & 0.00 & 27000 & 1.55\\
G15 & 5.30 & 0.00 & 31000 & 1.55\\
D13 & 2.20 & -0.45 & 18100 & 1.32\\
C10 & 1.45 & -1.00 & 13700 & 1.02\\
D10 & 2.20 & -1.00 & 18000 & 1.03\\
E10 & 3.10 & -1.00 & 22200 & 1.04\\
D07 & 2.20 & -1.60 & 17900 & 0.71\\
\hline 
\end{tabular}	
\label{table:simulations}	
\end{table}

\section{The curvature method}\label{sec:curvature}
The definition of curvature ($\kappa$), as used by \citet{Becker11}, is the following :
 \begin{equation} \label{c}
\kappa=\frac{F''}{[1+(F')^{2}]^{3/2}}   \ ,
\end{equation}
with the first and second derivatives of the flux ($F'$, $F''$) taken with respect to wavelength or relative velocity.
The advantage of this statistic is that, as demonstrated in \citet{Becker11}, it is quite sensitive to the IGM temperature but does not require the forest to be decomposed into individual lines. In this way the systematic errors are minimised if this analysis is applied to high resolution and high S/N spectra. Its calculation is relatively simple and can be computed using a single  b-spline fit directly to large regions of forest spectra. This statistic incorporates the temperature information from all lines, using more of the available information, as opposed to line-fitting which relies on selecting lines that are dominated by thermal broadening.    
If calibrated and interpreted using synthetic spectra, obtained from cosmological simulations, the curvature represents a powerful tool to measure the temperature of the IGM gas, $T(\bar{\Delta})$, at the characteristic overdensities ($\bar{\Delta}$) of the Lyman-$\alpha$ forest at different redshifts.

However, at low redshifts ($z\lesssim3$), the IGM gas tends to show characteristic overdensities ($\bar{\Delta}$) much higher than the mean density of the IGM ($\bar{\rho}$). Estimating the temperature at the mean density of the IGM ($T_{0}$) is then not straightforward. In fact, in the approximation of a gas collected into non-overlapping  clumps  of uniform density that have the same extent in redshift space as they have in real space, the Ly$\alpha$ optical depth at a given overdensity ($\Delta$) will scale as: 
 \begin{equation} \label{D}
\tau(\Delta) \propto (1+z)^{4.5} \Gamma^{-1} T_{0}^{-0.7}\Delta^{2-0.7(1-\gamma)}  \ ,
\end{equation}
where $\Gamma$ is the H I photoionisation rate and $T_{0}$ and $\gamma$ are the parameters that describe the thermal state of the IGM at redshift $z$ in Eq.1 (\citealt{Weinberg97}). In general, if we assume that the forest will be sensitive to overdensities that produce a Lyman-$\alpha$ optical depth $\tau(\Delta)\simeq 1$, it is then clear from Eq. 3 that these characteristic overdensities will vary  depending on the  redshift. At high redshift the forest will trace gas near  the mean density while in the redshift range of interest here ($z\lesssim3$) the absorption will be coming from densities increasingly above the mean. As a consequence, the translation of the $T(\bar{\Delta})$ measurements at the characteristic overdensities into the temperature at the mean density becomes increasingly dependent on the value of the slope of the temperature--density relation (Eq. 1). Due to the uncertainties related to a poorly constrained parameter $\gamma$, a degeneracy is introduced between $\gamma$ and the final results for $T_{0}$ that can be overcome only with a more precise measurement of the T--$\rho$ relation.

In this work we do not attempt to constrain the  full T--$\rho$ relation because that would require a simultaneous estimation of both $T_{0}$ and $\gamma$. Instead, following consistently the steps of the previous analysis of \citet{Becker11}, we establish empirically the characteristic overdensities ($\bar{\Delta}$) and obtain the corresponding temperatures from the curvature measurements. We define the characteristic overdensity traced by the Ly-$\alpha$ forest for each redshift as that overdensity at which $T(\bar{\Delta})$ is a one-to-one function of the mean absolute curvature, regardless of $\gamma$ (see Section 6).  We then recover the temperature at the mean density ($T_{0}$) from the temperature $T(\bar{\Delta})$ using Eq. 1 with a range of values of $\gamma$ (see Section 7).
\\

We can summarize our analysis in 3 main steps:  
  
\begin{itemize}
\item{ Data analysis (Section 5)}: from the selected sample of Lyman-$\alpha$ forest spectra we compute the curvature ($\kappa$) in the range of  $z\simeq 1.5-3.0$. From the observational spectra we also obtain measurements of the effective optical depth that we use to calibrate the simulations. 
 \item {Simulations analysis (Section 6)}: we calibrate the simulation snapshots at different redshifts in order to match the observational data. We obtain the curvature measurements from the synthetic spectra following the same procedure that we used for the real data and we determine the characteristic overdensities ($\bar{\Delta}$) empirically, finding  for each redshift the overdensity at which $T(\bar{\Delta})$  is a one-to-one function of $\log(\langle|\kappa|\rangle)$ regardless of $\gamma$.
 \item{Final temperature measurements (Section 7)}: we determine the $T(\bar{\Delta})$ corresponding to the observed curvature measurements by interpolating the $T(\bar{\Delta})$-$\log(\langle|\kappa|\rangle)$ relationship in the simulations to the values of $\log(\langle|\kappa|\rangle)$ from the observational data.
\end{itemize}

\section{Data analysis}\label{sec:analysis}
To directly match the box size of the simulated spectra, we compute the curvature statistic on sections of $10h^{-1}$ Mpc (comoving distance) of ``metal free''  Lyman-$\alpha$ forest regions in our quasar spectra. Metals lines are, in fact, a potentially serious source of systematic errors in any measure of the absorption features of the Ly-$\alpha$ forest. These lines tend to show individual components significantly narrower than the Lyman-$\alpha$ ones ($b\lesssim 15$ km$s^{-1}$) and, if included in the calculation, the curvature measurements will be biased towards high values. As a consequence, the temperature obtained will be much lower. For these reasons  we need to ``clean'' our spectra by adopting a comprehensive metal masking procedure (see Section 5.2).

However,  not only metals can affect our analysis and, even if  effectively masked from contaminant lines, the direct calculation of the curvature on observed spectra can be affected by other sources of uncertainties, particularly, noise and continuum errors.  To be as much as possible consistent with the previous work of \citet{Becker11} we adopted the same strategies to reduce these potential systematic errors.

\begin{figure} 
\centering
\includegraphics[width=0.5\textwidth]{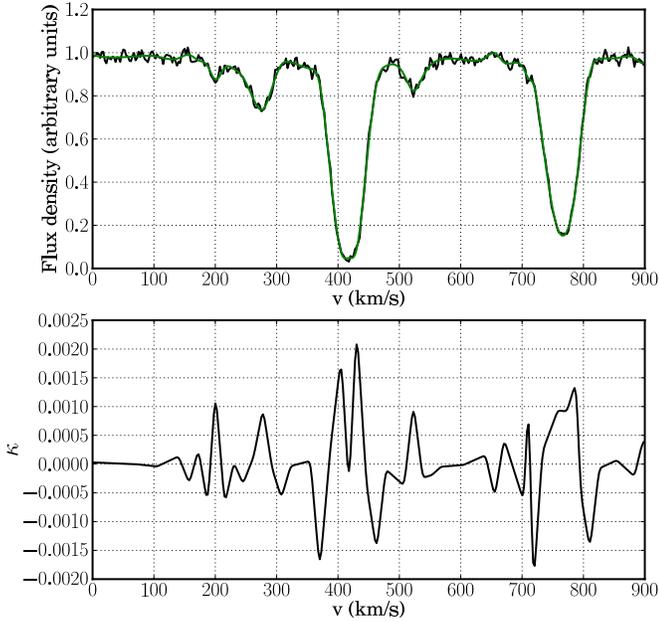} 
\caption{\small  Curvature calculation example for one section of Ly-$\alpha$ forest. Top panel: b-spline fit (green line) of a section of $10h^{-1}$ Mpc of normalized real spectrum. Bottom panel: the curvature statistic computed from the fit as defined in Eq. 2.}
\label{fig:kex}
\end{figure}

\emph{Noise}: If applied directly  to high resolution and high S/N spectra, the curvature measurements will be dominated by the noise in the flux spectra. To avoid this problem we fit a cubic b-spline to the flux and we then compute the curvature from the fit. In Figure \ref{fig:kex}, top panel, is shown a section of normalized Lyman-$\alpha$ spectra in which the solid green line is the b-spline fit from which we obtain the curvature. For consistency, we adopt the same specifics of the fitting routine of the previous work of \citet{Becker11}. We then use an adaptive fit with break points that are iteratively added, from an initial separation of 50 km $s^{-1}$, where the fit is poor. The iterations proceed until the spacing between break points reach a minimum value or the fit converges. With this technique we are able to reduce the sensitivity of the curvature to the amount of noise in the spectrum as we can test using the simulations (see section 6.2). 

\emph{Continuum}: Eq. 2 shows a dependence of the curvature on the amplitude of the flux, which in turn is dependent on the accuracy with which the unabsorbed quasar continuum can be estimated. The difficulty in determining the correct continuum level in the Lyman-$\alpha$ region can then constitute a source of uncertainties. To circumvent this issue we ``re-normalized" each $10h^{-1}$ Mpc section of data, dividing the flux of each section (already normalized by the longer-range fit of the continuum) by the maximum value of the b-spline fit in that interval.  Computing the curvature from the re-normalized  flux, we remove a potential systematic error due to inconsistent placement of the continuum. While this error could be important at high redshifts, where the Lyman-$\alpha$ forest is denser, at $z\lesssim 3$ we do not expect a large correction. In Figure \ref{fig:kex}, bottom panel, is shown the value of the curvature computed from the b-spline fit of the re-normalized flux (applying Eq. 2) for a section of forest.

We next measure the mean absolute curvature $\langle|\kappa|\rangle$ for the ``valid" pixels of each section.  We consider valid all the pixels where the re-normalized b-spline fit ($F^{\rm R}$) falls in the range $0.1 \leq F^{\rm R} \leq 0.9$.  In this way we exclude both the saturated pixels, that do not contain any useful information, and the pixels with flux near the continuum. This upper limit is in fact adopted because the flux profile tends to be flatter near the continuum and, as consequence the curvature for these pixels is considerably more uncertain. This potential uncertainty is particularly important at low redshift  because increasing the mean flux also increases the number of  pixels near the continuum (\citealt{Faucher-Giguere08}). 
 
\subsection{Observed curvature and re-normalized optical depth}
\begin{figure} 
\centering
\includegraphics[width=0.45\textwidth]{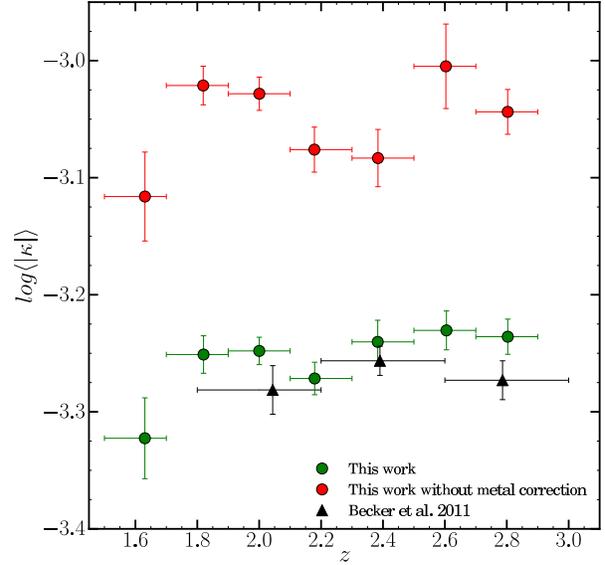} 
\caption{\small  Curvature measurements from the observational quasar spectra. The curvature values obtained in this work (green points) for redshift bins of $\Delta z=0.2$ are compared with curvature points from \citet{Becker11} with $\Delta z=0.4$ (black triangles). Horizontal error bars show the redshift range spanned by each bin. Vertical error bars in this work are 1$\sigma$ and have been obtained from a bootstrap technique using the curvature measurements within each bin. In Becker et al. the errors are 2$\sigma$, recovered using sets of artificial spectra. These errors from the simulations are in agreement with the direct bootstrap using the data from bins which contain a a large number of data points. Curvature measurements obtained in this work from spectra not masked for metals are also shown (red points).  }
\label{fig:kcomp}
\end{figure}

The final results for the curvature measurements from the real quasar spectra are shown in the green data points in Figure \ref{fig:kcomp}. In this plot the values of  $\langle|\kappa|\rangle$ obtained from all the $10h^{-1}$ Mpc sections of forest have been collected and averaged in redshift bins of $\Delta z =0.2$. The error bars show the $1\sigma$ uncertainty obtained with a bootstrap technique generated directly from the curvature measurement within each bin. In all the redshift bins, in fact, the mean absolute curvature values of a large number of sections ($N>100$) have been averaged and so the bootstrap can be considered an effective tool to recover the uncertainties. It is important to note that the smaller number of sections contributing to the lowest redshift bin ($1.5\leq z\leq 1.7$; see Figure \ref{fig:histo}), is reflected in a larger error bar.
For comparison are shown the results of the curvature from \citet{Becker11} (black triangles) for redshift bins of $\Delta z=0.4$. In the  common redshift range the results seem to be in general agreement even if our values appear shifted slightly towards higher curvatures. Taking into consideration the fact that each point cannot be considered independent from the neighbours, this shift between the results from the two different data samples is not unexpected and may also reflect differences in the signal-to- noise ratio between the samples. At this stage we do not identify any obvious strong departure from a smooth trend in $\langle|\kappa|\rangle$ as a function of redshift.\\

\begin{figure} 
\centering
\includegraphics[width=0.45\textwidth]{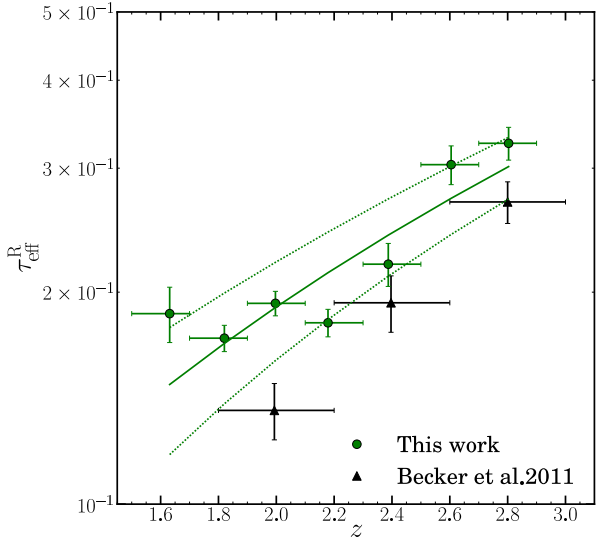} 
\caption{\small Effective re-normalized optical depth ($\tau_{\rm eff}^{\rm R}$) from our quasar spectra  (green points) compared with the results from \citet{Becker11} (black triangles). Vertical error bars are 1$\sigma$ bootstrap uncertainties for our points and $2\sigma$ for the previous work, while the horizontal bars show the redshift range spanned by each bin. The solid line represents the least square fit from our measurements while the dotted lines show the $\pm1\sigma$ confidence interval on the fit. The measurements have been obtained directly from quasar spectra after correcting for metal absorption.}
\label{fig:taucomp}
\end{figure}

From  each section we also extract the mean re-normalized flux ($F^{\rm R}$) that  we use to estimate the re-normalized effective optical depth ($\tau_{\rm eff}^{\rm R}=-ln\langle F^{\rm R}\rangle$) needed for the calibration of the simulations (see Section 6.1). In Figure \ref{fig:taucomp} we plot the $\tau_{\rm eff}^{\rm R}$obtained in this work for redshift bins of $\Delta z= 0.2$ (green data points) compared with the results of \citet{Becker11} for bins of $\Delta z= 0.4$ (black triangles). Vertical error bars are 1$\sigma$ bootstrap uncertainties for our points and $2\sigma$ for \citet{Becker11}. For simplicity we fitted our data with a unique power law  ($\tau=A(1+z)^\alpha$) because we do not expect that a possible small  variation of the slope as a function of redshift will have a relevant effect in the final temperature measurements. Comparing the least square fit computed from our measurements (green solid line) with the re-normalized effective optical depth of Becker et al. it is evident that there is a systematic difference between the two samples, increasing at lower redshifts: our $\tau_{\rm eff}^{\rm R}$ values are $\sim 10\% $ higher compared with the previous measurements and, even if the black triangles tend to return inside the $\pm1\sigma$ confidence interval on the fit (green dotted lines) for $z \gtrsim 2.6$, the results are not in close agreement. However, the main quantities of interest in this paper (i.e.~the curvature, from which the IGM temperature estimates are calculated) derive from a comparison of real and simulated spectra which have been re-normalized in the same way, so we expect that they will not depend strongly on the estimation of the continuum like the two sets of $\tau^{\rm R}_{\rm eff}$ results in Fig.~5 do (we compare the effective optical depth prior the continuum re-normalization from the simulations calibrated with the $\tau^{\rm R}_{\rm eff}$ results in Section 6.1.2). The higher values shown by our re-normalized effective optical depth reflect the variance  expected between different samples: a systematic scaling, of the order of the error bar sizes, between our results and the previous ones may not be unexpected due to the non-independence of the data points within each set.                     
 These differences will reflect  different characteristic overdensities probed by the forest (see Section 6.3). Fortunately, the correct calibration between  temperature and curvature measurements will wash out this effect, allowing consistent temperature calibration as a function of redshift (see Section 7 and Appendix A). 

Even if the scaling between the datasets could be smoothed, considering the fact that, according to \citet{Rollinde13}, the bootstrap errors computed from sections of 10 $h^{-1}$Mpc (and then $\lesssim 25 \AA$)  could underestimate the variance, another possible cause could be differences in the metal masking procedures of the two studies. In the next section we explain and test our metal masking technique, showing how our results do not seem to imply a strong bias due to contamination from unidentified metal lines.

\subsection{Metal correction}
Metal lines can be a  serious source of systematic uncertainties for both the measures of the re-normalized flux ($F^{\rm R}$) and the curvature.  While in the first case it is  possible to choose between a statistical (\citealt{Tytler04}; \citealt{Kirkman05}) and a direct (\citealt{Schaye03}) estimation of the metal absorption, for the curvature it is necessary to directly identify and mask individual metals lines in the Lyman-$\alpha$ forest. Removing these features accurately is particularly important for redshifts $z\lesssim 3$, where there are fewer Ly$\alpha$ lines and potentially their presence could affect significantly the results. We therefore choose to identify metal lines proceeding in two steps: an ``automatic"  masking procedure followed by a manual refinement. First, we use well-known pairs of strong metal-line transitions to find all the obvious metal absorbing redshifts in the spectrum. We then classify each absorber as of high (e.g. C IV, Si IV) or low (e.g. Mg II, Fe II) ionisation and strong or weak absorption, and we evaluate the width of its velocity structure. To avoid contamination, we mask all the regions in the forest that could plausibly contain common metal transitions of the same type, at the same redshift and within the same velocity width of these systems. The next step is to double-check the forest spectra by eye, searching for remaining unidentified narrow lines (which may be metals) and other contaminants (like damped Lyman-$\alpha$ systems or corrupted chunks of data). Acknowledging that this procedure is in a certain way subjective, at this stage  we try to mask any feature with very narrow components or sharp edges to be conservative.

\begin{figure} 
\centering
\includegraphics[width=0.45\textwidth]{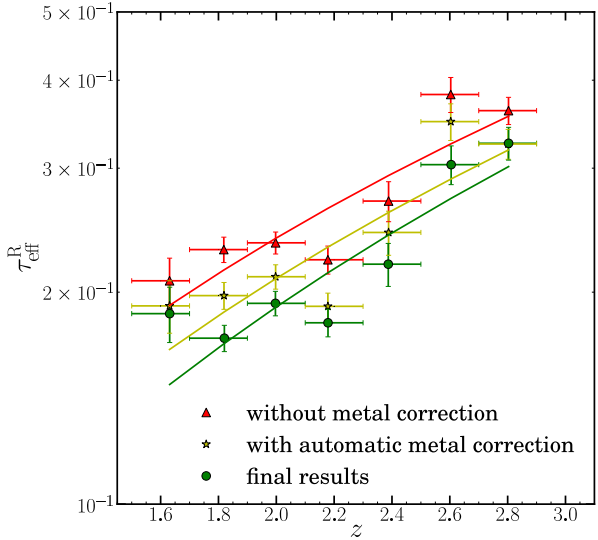} 
\caption{\small Comparison of our final measurements of the re-normalized effective optical depth ($\tau_{\rm eff}^{\rm R}$) (green points) with the values without applying any metal correction to the spectra (red triangles) and with only the first, automatic correction (yellow stars). Solid lines are the least square fits to the corresponding points, while vertical bars represent the $1\sigma$ bootstrap errors. The metal absorption has been estimated following the procedure described in the text.} 
\label{fig:taumetals}
\end{figure}

In Figure \ref{fig:taumetals} is presented the correction for the metal line absorption on the re-normalized effective optical depth measurements; from the raw spectra (red triangles), to the spectra treated with the first ``automatic" correction (yellow stars), to the final results double-checked by eye (green points). For all the three cases we show the least square fit (solid lines of corresponding colors) and the 1$\sigma$ vertical error bars. In Table \ref{table:metalcorr} are reported the numerical values for our metal absorption compared with previous results of \citet{Schaye03} and \citet{Kirkman05} used in the effective optical depth measurements of \citet{Faucher-Giguere08}. The relative metal correction to $\tau_{\rm eff}^{\rm R}$ decreases with increasing redshift, as expected  if the IGM is monotonically enriched with time (\citealt{Faucher-Giguere08}) and in general is consistent with the previous results. In their work in fact, Faucher-Gigu\`{e}re et al. evaluated the relative percentages of metal absorption in their measurements of $\tau_{\rm eff}$ when applying two different corrections: the one obtained with the direct identification and masking method by \citet{Schaye03}, and the statistical estimate of \citet{Kirkman05} in which they used measurements of the amount of metals redwards the Ly$\alpha$ emission line. At each redshift, \citet{Faucher-Giguere08} found good agreement between the estimates of their effective optical depth based on the two methods of removing metals.  In  individual redshift bins, applying our final relative metal absorption percentages, we obtain a $\tau_{\rm eff}^{\rm R}$ that agrees well within $1\sigma$  with the ones obtained after applying the corrections of these previous results, encouraging confidence that our metal correction is accurate to the level of our statistical error bar. However, our corrections are overall systematically larger than the previous ones. If we have been too conservative in removing potentially metal-contaminated portions of spectra  in our second ``by-eye" step, this will bias the effective optical depth to lower values.
    
In Figure \ref{fig:kcomp}  is also shown the effect of our metal correction on the curvature measurements: red points are curvature values obtained from the raw spectra while the green points are the final measurements from masked sections, with vertical bars being the $1\sigma$ error. Metal contamination has important effects on the curvature measurements: after the correction, in fact, the curvature measurements decrease between $\sim 30\%-40\%$ at each redshift even if the relative differences among redshift bins seem to be maintained. The potential effects of an inaccurate metal correction on the final temperature measurements will be considered in Section 7.

\begin{table} 
\caption{\small Metal absorption correction to the raw measurements of $\tau_{\rm eff}^{\rm R}$. For each redshift (column 1) is reported the percentage metal absorption correction obtained in this work in the first, `automatic' mask described in the text (column 2) and in the refinement bye eye (column 3). For comparison, in the overlapping redshift range are presented the results of \citet{Faucher-Giguere08} obtained applying the direct metal correction of \citet{Schaye03} (column 4) and the statistical one of \citet{Kirkman05} (column 5).} 
\centering \begin{tabular}{c c c c c } 
\hline\hline
z\ &  automatic corr.\ & final corr.\ & Schaye corr.\ & Kirkman corr.\ \\
\hline 
1.6 & $13.8\%$ & $22.9\%$ & n.a. & n.a.\\
1.8 & $13.2\%$ & $21.5\%$ & n.a. & n.a.\\
2.0 & $12.6\%$ & $20.1\%$ & $13.0\%$ & $21.0\%$\\
2.2 & $12.0\%$ & $18.8\%$ & $12.3\%$ & $16.0\%$\\
2.4 & $11.5\%$ & $17.5\%$ & $11.4\%$ & $12.6\%$\\
2.6 & $11.0\%$ & $16.3\%$ & $10.4\% $& $10.4\%$\\
2.8 & $10.5\%$ & $15.1\%$ & $9.7\%$ & $7.8\%$\\
3.0 & $8.8\%$ & $14.0\%$ & $9.0\%$ & $6.0\%$\\

\hline 
\end{tabular}	
\label{table:metalcorr}	
\end{table}

\subsection{Proximity region}
\begin{figure} 
\centering
\includegraphics[width=0.45\textwidth]{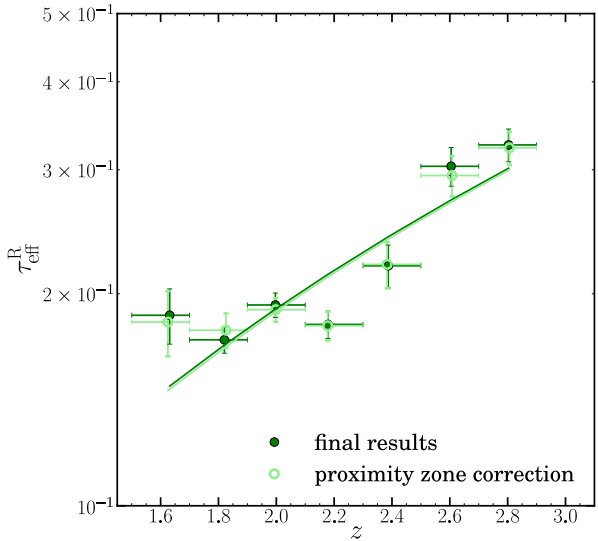} 
\caption{\small  Effect of masking the proximity region on the re-normalized effective optical depth. The final results for $\tau_{\rm eff}^{\rm R}$ without the masking of the proximity zones are shown as green points and those with this correction are shown as light green circles. Solid lines represent the least square fit of the data and vertical error bars are the $1\sigma$ statistical uncertainties. }
\label{fig:tauprox}
\end{figure}

A final possible contaminant in the measurements of the effective optical depth is the inclusion in the analysis of the QSO proximity regions.
The so-called ``proximity regions'' are the zones near enough to the quasars to be subjected to the local influence of its UV radiation field. These areas may be expected to show lower Ly-$\alpha$ absorption with respect to the cosmic mean due to the high degree of ionisation. To understand if the proximity effect can bias the final estimates of $\tau_{\rm eff}^{\rm R}$, we compare our results with the measurements obtained after masking the chunks of spectra that are potentially effected by the quasar radiation. Typically the ionizing UV flux of a bright quasar is thought to affect regions of $\lesssim 10 $ proper Mpc along its own line of sight (e.g. \citealt{Scott00}; \citealt{Worseck06}). To be conservative, we masked the 25 proper Mpc nearest to each quasar Ly-$\alpha$ and Ly-$\beta$ emission lines and re-computed $\tau_{\rm eff}^{\rm R}$. The final comparison is presented in Figure \ref{fig:tauprox}: masking the proximity regions does not have any significant effect on $\tau_{\rm eff}^{\rm R}$. In fact, the results obtained excluding these zones (light green circles) closely match the results inferred without this correction (green points), well within the 1$\sigma$ error bars. We then do not expect that the inclusion in our analysis of the QSO proximity regions will affect significantly the temperature measurements.

\section{Simulations Analysis}
To extract temperature constraints from our measurements of the curvature we need to interpret our observational results using simulated spectra, accurately calibrated to match the real data conditions. In this Section we explain how we calibrate and analyse the synthetic spectra to find the connection between curvature measurements and temperature at the characteristic overdensities. We will use these results in Section 7 where we will interpolate the $T(\bar{\Delta})$--$log\langle|\kappa|\rangle$ relationship  to the value of  $\log\langle|\kappa|\rangle$ from the observational data to obtain our final temperature measurements.
\subsection{The calibration}
To ensure a correct comparison between simulation and observational data we calibrate our synthetic spectra to match the spectral resolution and the pixel size of the real spectra. We adjust the simulated, re-normalized effective optical depth ($\tau_{\rm eff}^{\rm R}$) to the one extracted directly from the observational results (see Section 5.1) and we add to the synthetic spectra the same level of noise recovered from our sample.
\subsubsection{Addition of noise}
\begin{figure} 
\centering
\includegraphics[width=0.5\textwidth]{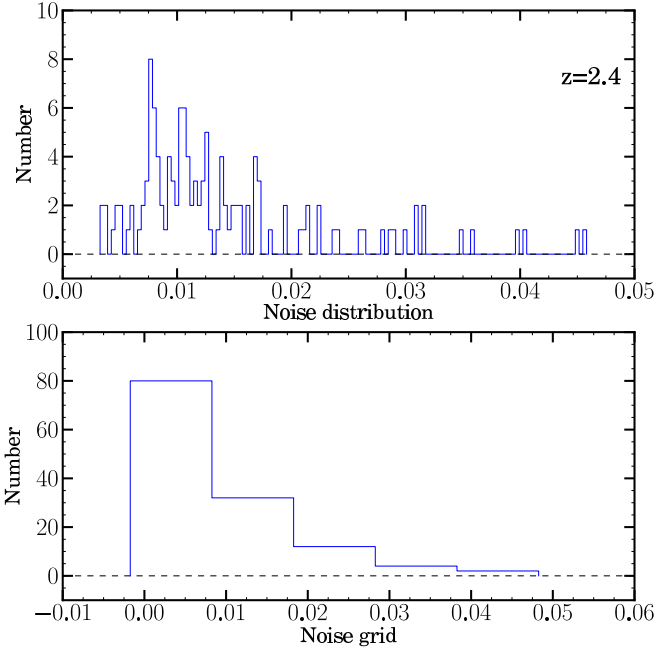} 
\caption{\small  Top panel: an example of noise distribution for a  $\Delta z=0.2$ redshift bin of real quasar spectra (in this case  one with $z_{mean}=2.4$). On the x-axis of the histogram is presented the mean noise per section of 10$h^{-1}$ Mpc while on the y-axis is shown the number of sections contributing to the particular bin. Bottom panel: the same distribution presented in the top panel but simplified, collecting the data in a noise grid of $\Delta\sigma=0.01$.}
\label{fig:noise}
\end{figure}

To add the noise to the synthetic spectra we proceed in three steps: first,  we obtain the distributions of the mean noise corresponding to the  10$h^{-1}$Mpc sections  of the quasar spectra contributing to each redshift bin. As shown in Figure \ref{fig:noise} (top panel) these distributions can be  complex and so to save computational time we simplify them by extracting grids of noise values with a separation of $\Delta\sigma=0.01$ and weights rescaled proportionally to the original distribution (Figure \ref{fig:noise} bottom panel). At each redshift the noise is finally added at the same levels of the corresponding noise distribution and the quantities computed from the synthetic spectra, with different levels of noise, are averaged with the weights of the respective noise grid.
\subsubsection{Recovered optical depth }
\begin{figure} 
\centering
\includegraphics[width=0.5\textwidth]{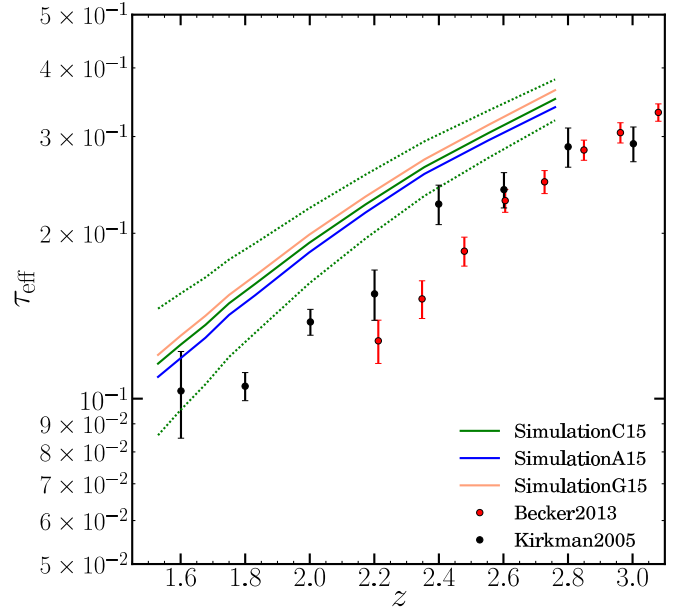} 
\caption{\small The effective optical depth, prior to the re-normalization correction, recovered from the simulations which matches the re-normalized effective optical depth, $\tau_{\rm eff}^{\rm R}$, of our real spectra.  The effective optical depth for three different simulations is shown: A15 (blue solid line), G15 (pink solid line) and C15 (green solid line). The recovered $\tau_{\rm eff}$ for the remaining simulations in Table 2 are not reported for clarity but they lie in between the trends of A15 and G15. The spread in values for different thermal histories is, in fact, small compared with the 1$\sigma$ uncertainties about the trend of each of the effective optical depths (green dotted lines for simulation C15). Our results are compared with the effective optical depths of \citet{Becker12} (red points) and \citet{Kirkman05} (black points). }
\label{fig:taurecov}
\end{figure}

The simulated spectra are scaled to match the re-normalized effective optical depth, $\tau_{\rm eff}^{\rm R}$, of  the real spectra. These can then be used to recover the corresponding effective optical depth ($\tau_{\rm eff}$; prior to the continuum re-normalization).
 In fact, the re-normalized effective optical depth cannot be compared directly  with the results from the literature and to do so we need to compute the mean flux and then $\tau_{\rm eff}$ ($\tau_{\rm eff}= -ln\langle F\rangle$) from the synthetic spectra using the same procedure applied previously to the real spectra (see Section 5) but without the re-normalization. In Figure \ref{fig:taurecov} is shown the trend of the recovered effective optical depth for three different simulations, A15, G15 and C15 in Table \ref{table:simulations}.  For clarity  we do not plot the curves for the remaining simulations but they lie in between the curves of simulations A15 and G15. Depending on the different thermal histories, the recovered effective optical depths vary slightly but the separation of these values is small compared with the uncertainties about the trend  (e.g. green dotted lines referred to the simulation C15).

In Figure \ref{fig:taurecov} we also compare our results with the previous studies of \citet{Becker12}, and \citet{Kirkman05}. The results of \citet{Becker12}, that for $z\lesssim2.5$ have been scaled to the \citet{Faucher-Giguere08} measurements, are significantly shifted toward lower $\tau_{\rm eff}$, presenting a better agreement with Kirkman et al.. For $z< 2.2$ the effective optical depth of Kirkman et al. still shows values $\sim 30\%$ lower than ours.  In this case, again, part of the difference between the results could be explained by the non-independence of the data points within each set.  Such an offset could also be boosted by a possible selection effect: the lines of sight used in this work were taken from the UVES archive and, as such, may contain a higher proportion of damped Lyman-$\alpha$ systems; even if these systems have been masked out of our analysis, their presence will increase the clustering of the forest around them and so our sample will have higher effective optical depth as a consequence. The simulated $\tau_{\rm eff}$ presented in Figure \ref{fig:taurecov} were obtained by matching the observed $\tau_{\rm eff}^{\rm R}$ (see Figure \ref{fig:taucomp}) and do not represent one of the main results of this work, so we did not investigate further possible selection effects driven by our UVES sample. Being aware of this possibility, we decided to maintain the consistency between our curvature measurements and the simulations used to infer the temperature values, calibrating the simulated spectra with the effective optical depth obtained from our sample (see Appendix A). In the comparison between our results and the previous ones of \citet{Becker11}, the effect of a calibration with an higher $\tau_{\rm eff}$ will manifest itself as a shift towards lower values in the characteristic overdensities traced by the Lyman-$\alpha$ forest at the same redshift (as we will see in Section 6.3).

\subsection{ The curvature from the simulations}
\begin{figure} 
\centering
\includegraphics[width=0.5\textwidth]{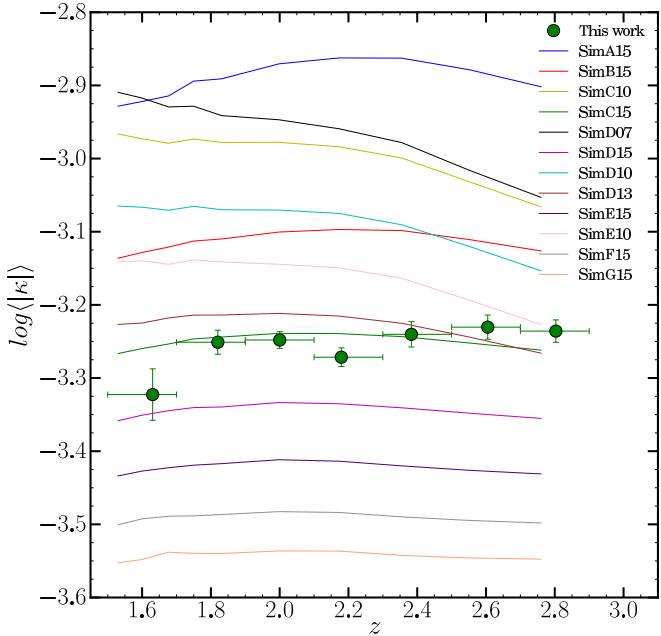} 
\caption{\small Curvature measurements: points with vertical error bars ($1\sigma$ uncertainty) are for the real data and are compared with the curvature obtained from simulations with different thermal histories calibrated with the same spectral resolution, noise and effective optical depth of the observed spectra at each redshift.}
\label{fig:kfavorite}
\end{figure}

Once the simulations have been calibrated we can measure the curvature on the synthetic spectra using the same method that we used for the observed data (see Section 5).
In Figure \ref{fig:kfavorite} are plotted the values of $\log\langle|\kappa|\rangle$ obtained from our set of simulations in the same redshift range and with the same spectral resolution, effective optical depth and mix of noise levels of the real spectra. Different lines correspond to different simulations in which the thermal state parameters are changed. We can preliminarily compare our data points with the simulations, noticing that the simulation that has the values of the curvature close to the real observations is C15, which assumes the fiducial parameter  $\gamma=1.54$ at redshift $z$=3. 
The variation in the properties of the real data alters the trend of the simulated curvature to be a slightly non-smooth function of redshift.

\begin{figure} 
\centering
\includegraphics[width=0.45\textwidth]{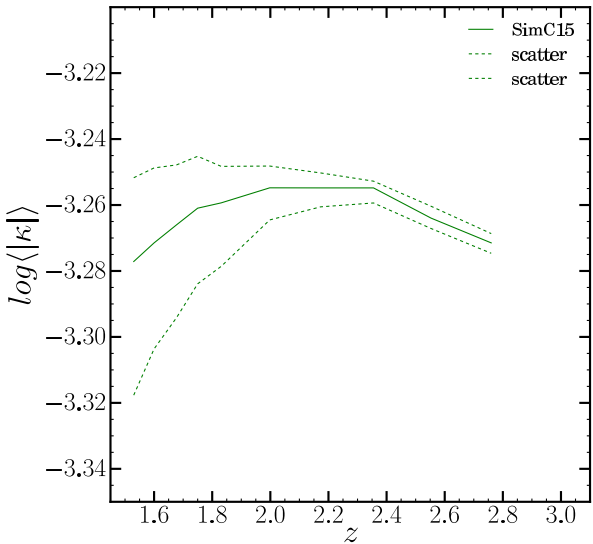} 
\caption{\small Dependence of the simulated $\log\langle|\kappa|\rangle$ on the effective optical depth with which the simulations have been calibrated. The curvature recovered using the thermal history C15 (green solid line) is reported with the 1$\sigma$ uncertainties about the trend (green dotted lines) corresponding to the $1\sigma$ uncertainties about the fit of the observed  $\tau_{\rm eff}^{R}$ in Figure 5. The variation generated in the curvature is about $10\%$ for $z=1.5$ corresponding to a scatter of $\pm2-3\times 10^{3}$ K in the temperature calibration, and decreases at higher redshift.  }
\label{fig:kerror}
\end{figure}

As expected, the curvature values are sensitive to changes in the effective optical depth: as shown in Figure \ref{fig:kerror} for the fiducial simulation C15, the $1\sigma$ uncertainty about the fit of the observed $\tau_{\rm eff}^{\rm R}$ (see Figures 5) is in fact reflected in a scatter about the simulated curvature of about $10\% $  at redshift $z\sim 1.5$, decreasing at higher redshifts. The next section shows how this dependence of the simulated curvature on the matched effective optical depth will imply differences in the recovered characteristic overdensities between our work and \citet{Becker11}.
    

\subsection{ The characteristic overdensities}
The final aim of this work is to use the curvature measurements to infer information about the thermal state of the IGM, but this property will depend on the density of the gas. The Lyman-$\alpha$ forest, and so the curvature obtained from it, in fact does not always trace the gas at the mean density but, instead, at low redshift ($z\lesssim3$) the forest lines will typically arise from densities that are increasingly above the mean. The degeneracy between $T_{0}$  and $\gamma$ in the temperature--density relation (Eq.1) will therefore be significant. For this reason in our work we are not constraining both these parameters but we use the curvature to obtain the temperature at those characteristic overdensities ($\bar{\Delta}$) probed by the forest that will not depend on the particular value of  $\gamma$. We can in this way associate uniquely our curvature values to the temperature at these characteristic overdensities, keeping  in mind that the observed values of $\kappa$ will represent anyway an average  over a range of densities.
\subsubsection{The method}
We determine the characteristic overdensities empirically, finding for each redshift the overdensities at which $T(\bar{\Delta})$ is a one-to-one function  of $\log\langle|\kappa|\rangle$ regardless of $\gamma$. The method is explained in Figure 12: for each simulation type we plot the values of $T(\Delta)$ versus $\log\langle|\kappa|\rangle$, corresponding to the points with different colours, and we fit the distribution with a simple power law. We change the value of the overdensity $\Delta$ until we find the one ($\bar{\Delta}$) for which all the points from the different simulations (with different thermal histories and $\gamma$ parameters) lie on the same curve and minimize the $\chi^{2}$. The final $T(\bar{\Delta})$ of our real data (see section 7) will be determined by interpolating the $T(\bar{\Delta})$--$\log\langle|\kappa|\rangle$ relationship in the simulations to the value of $\log\langle|\kappa|\rangle$ computed directly from the real spectra.
\subsubsection{The results}
The characteristic overdensities for the redshifts of our data points are reported in Table \ref{table:results}, while in Figure \ref{fig:overdensity} is shown the evolution of $\bar{\Delta}$ as a function of redshift: as expected, at decreasing redshifts the characteristic overdensity at which the Lyman-$\alpha$ forest is sensitive increases. Note that Figure \ref{fig:overdensity} also shows that while the addition of noise in the synthetic spectra for $z\lesssim 2.2$ has the effect of decreasing the values of the characteristic overdensities, that tendency is inverted for higher redshifts where the noise shifts the overdensities slightly towards higher values with respect to the noise-free results. In Figure \ref{fig:overdensity} is also presented a comparison between the overdensities founded in this work and the ones obtained in \citet{Becker11} in their analysis with the addition of noise. Even if the two trends are similar, the difference in values of the characteristic overdensities at each redshift is significant ($\sim 25\%$ at  $z\sim 3$ and increasing towards lower redshift). Because we used the same set of thermal histories and a consistent method of analysis with respect to the previous work, the reason for this discrepancy lies in the different data samples: in fact, the effective optical depth observed in our sample is higher than the one recorded by \citet{Becker11} (see Section 5.1). As we have seen in Section 6.2, the simulated curvature is sensitive to the effective optical depth with which the synthetic spectra have been calibrated and this is reflected in the values of the characteristic overdensities.  It is then reasonable that for higher effective optical depths at a particular redshift we observe lower overdensities because we are tracing a denser universe and the Lyman-$\alpha$ forest will arise in overdensities closer to the mean density.
\begin{figure} 
\centering
\includegraphics[width=0.4\textwidth]{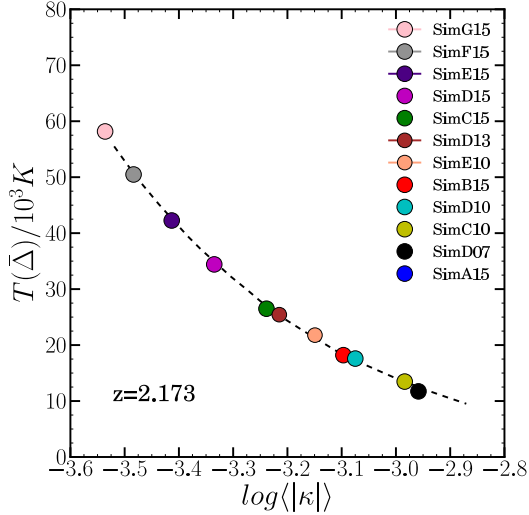} 
\caption{\small Example of the one-to-one function between $\log\langle|\kappa|\rangle$ and  temperature obtained for a characteristic overdensity ($\bar{\Delta}=3.7$ at redshift $z=2.173$),   Different colors correspond to different simulations. At each redshift we find the characteristic overdensity, $\Delta=\bar{\Delta}$, for which the relationship between $T(\bar{\Delta})$-- $\log\langle|\kappa|\rangle$ does not depend on the choice of a particular thermal history or $\gamma$ parameter.}
\end{figure}     
 
\begin{figure} 
\centering
\includegraphics[width=0.4\textwidth]{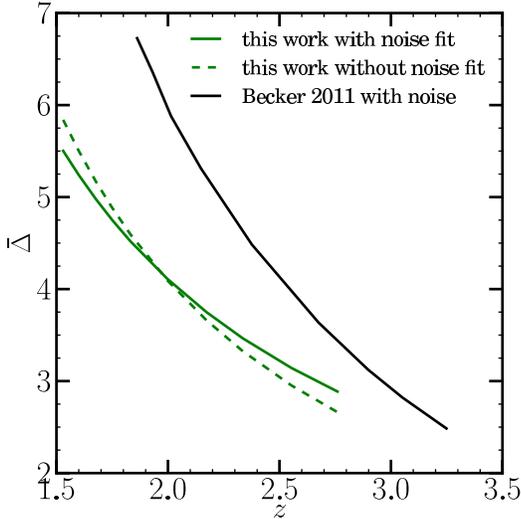} 
\caption{\small Evolution as a function of redshift of the characteristic overdensities obtained in this work with the addition of noise in the synthetic spectra (green solid line) and with noise-free simulations (green dotted line). As expected, the characteristic overdensities traced by the Lyman-$\alpha$ forest increase toward lower redshift. For comparison, the result for the characteristic overdensities from the previous work of \citet{Becker11} (black solid line) is also presented. Our overdensities are lower and this can be associated with the higher effective optical depth observed in our sample that was used to calibrate the simulations. }
\label{fig:overdensity}
\end{figure}     

\begin{table} 
\caption{\small Numerical values for the results of this work: the mean redshift of each data bin is reported (column 1) with the associated characteristic overdensity (column 2). Also shown are the temperature measurement with the associated 1$\sigma$ errors obtained for each data bin at the characteristic overdensity (column 3) and at the mean density under the assumption of two values of $\gamma$ (column 4 \& 5). Finally, the values of $\gamma$, recovered from the fiducial simulation C15, are presented (column 4).  } 
\centering \begin{tabular}{c c c c c c } 
\hline\hline
$z_{mean}$\ &  $\bar{\Delta}$\ & $T(\bar{\Delta})/10^{3}K$\ & $\gamma\sim 1.5$\ & $T_{0}^{\gamma\sim 1.5}/10^{3} K$\ & $T_{0}^{\gamma=1.3}/10^{3} K$\ \\
\hline 
1.63 & 5.13 & 33.74$\pm$3.31 & 1.583 & 13.00$\pm$1.27 & 20.66$\pm$2.03\\
1.82 & 4.55 & 27.75$\pm$1.19 & 1.577 & 11.79$\pm$0.51 & 17.61$\pm$0.76 \\
2.00 & 4.11 &  27.62$\pm$0.84 &1.572  & 12.60$\pm$0.38 & 18.08$\pm$0.55\\
2.18 & 3.74 & 29.20$\pm$1.06 &1.565 & 14.05$\pm$0.51 & 19.66$\pm$0.71\\
2.38 & 3.39 & 25.95$\pm$1.22 & 1.561 & 13.20$\pm$0.62 & 18.00$\pm$0.85\\
2.60 & 3.08 & 23.58$\pm$1.09 &1.554 &  12.77$\pm$0.59 & 16.83$\pm$0.78\\
2.80 & 2.84 & 22.67$\pm$0.89& 1.549& 12.90$\pm$0.51 & 16.58$\pm$0.65\\

\hline 
\end{tabular}
\label{table:results}		
\end{table}

\section{Temperature measurements}
  The selection of the characteristic overdensities, $\bar{\Delta}$, and the associated one-to-one function between temperature and curvature allows us to infer information about the temperature of the gas traced by the Lyman-$\alpha$ forest, $T(\bar{\Delta})$. This measurement is independent of the choice of the parameter $\gamma$ for the T--$\rho$ relation (Eq. 1) and for this reason will represent the main result of this work. We also translate our temperature measurements to values at the temperature at the mean density, $T_{0}$, for reasonable values of $\gamma$. In this Section we present our results and compare them with those of \citet{Becker11} at higher redshift. A broader discussion, taking into consideration theoretical predictions,  can be found in Section 8.
\subsection{ Temperature at the characteristic overdensities}
\begin{figure} 
\centering
\includegraphics[width=0.5\textwidth]{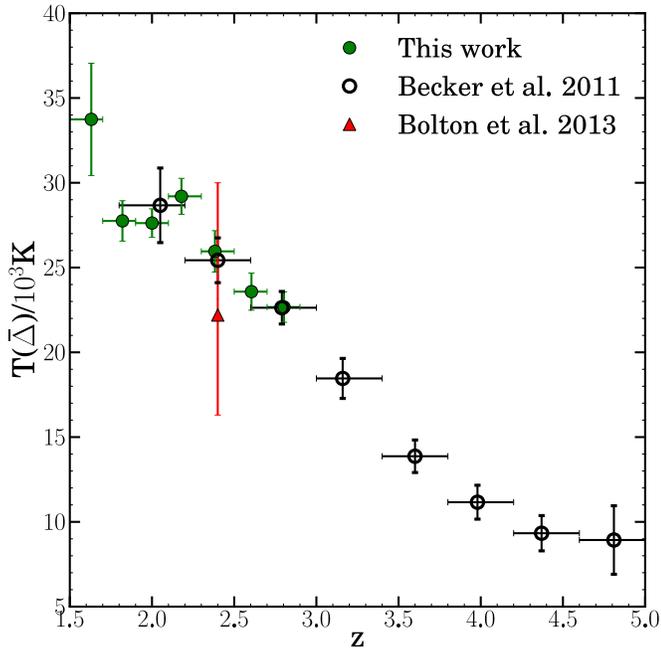} 
\caption{\small IGM temperature at the characteristic overdensities, $T(\bar{\Delta})$, as a function of redshift for this work (green points), for the line-fitting analysis of \citet{Bolton13} (red triangle) and for the previous work of \citet{Becker11} (black circles). Vertical error bars are $1\sigma$  for this work and $2\sigma$  for Becker et al.  and are estimated from statistical uncertainties in the curvature measurements.  }
\label{fig:TD}
\end{figure}

The main results of this work are presented in Figure \ref{fig:TD} where we plot the IGM temperature at the characteristic overdensities traced by the Ly-$\alpha$ forest as a function of redshift. The 1$\sigma$ errors are estimated from the propagation of the uncertainties in the curvature measurements. In fact, the uncertainties in the measured effective optical depth are reflected only in a small variation in the temperature measurements that falls well within the 1$\sigma$ uncertainties due to the errors in the curvature measurements. Our temperature measurements show good agreement with the previous work of \citet{Becker11} at higher redshifts where they overlap. This accord is particularly significant because we analysed a completely independent set of quasar spectra, obtained from a different instrument and telescope.

The curvature method in fact demonstrates self-consistency: the lower overdensities recorded from our sample are in fact compensated by our higher values of observed curvature. In this way, interpolating at each redshift the $T(\bar{\Delta})$--$\log\langle|\kappa|\rangle$ relationship in the simulations to the $\log\langle|\kappa|\rangle$ computed directly from the data, we obtained similar temperature values to the ones of \citet{Becker11} in the overlapping redshift range. Differences in the characteristic overdensities, $\bar{\Delta}$, at a particular redshift between the two studies will cause variation in the derived temperature at the mean density ($T_{0}$) because we will infer $T_{0}$ using the T--$\rho$ relation with the values of $\bar{\Delta} $. However, this effect will be modest and will cause disparity at the level of the 1$\sigma$ error bars of our values (see Section 7.2 and Appendix A). For comparison, in Figure \ref{fig:TD} we show the $z=2.4$ line-fitting result of \citet{Rudie12}, with their $T_{0}$ and $\gamma$ values recalibrated and translated to a $T(\bar{\Delta})$ value by \citet{Bolton13}. Even if the line-fitting method is characterized by much larger 1$\sigma$ error bars, it represents an independent technique and its agreement with our temperature values gives additional confidence in the results.

In general, the extension to lower redshifts ($z\lesssim1.9$) that our new results provide in Figure \ref{fig:TD} do not show any large, sudden decrease or increase in $T(\bar{\Delta})$ and can be considered broadly consistent with the trend of $T(\bar{\Delta})$ increasing towards lower redshift  of \citet{Becker11}. The increasing of $T(\bar{\Delta})$ with decreasing redshift is expected for a non-inverted temperature--density relation because, at lower $z$, the Lyman-$\alpha$ forest is tracing higher overdensities: denser regions are much more bounded against the cooling due to the adiabatic expansion and present higher recombination rates (and so more atoms for the photoheating process). We consider this expectation further in Section 8 after converting our $T(\bar{\Delta})$ measurements to $T_{0}$ ones, using a range of $\gamma $ values.

\subsection {Temperature at the mean density}
\begin{figure}

\centering

\begin{tabular}{cc}

\includegraphics[width=0.4\textwidth]{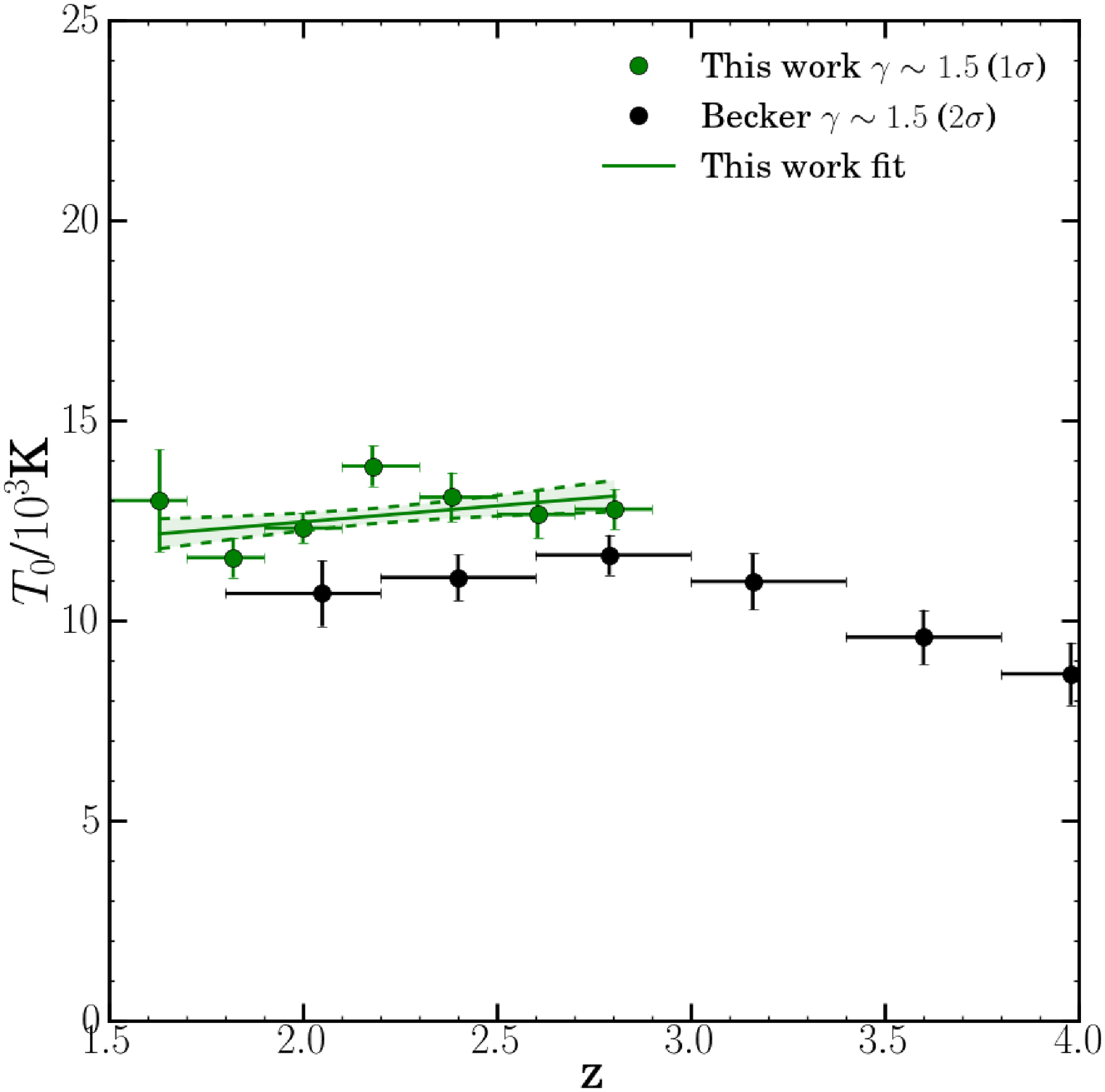}\\

\includegraphics[width=0.4\textwidth]{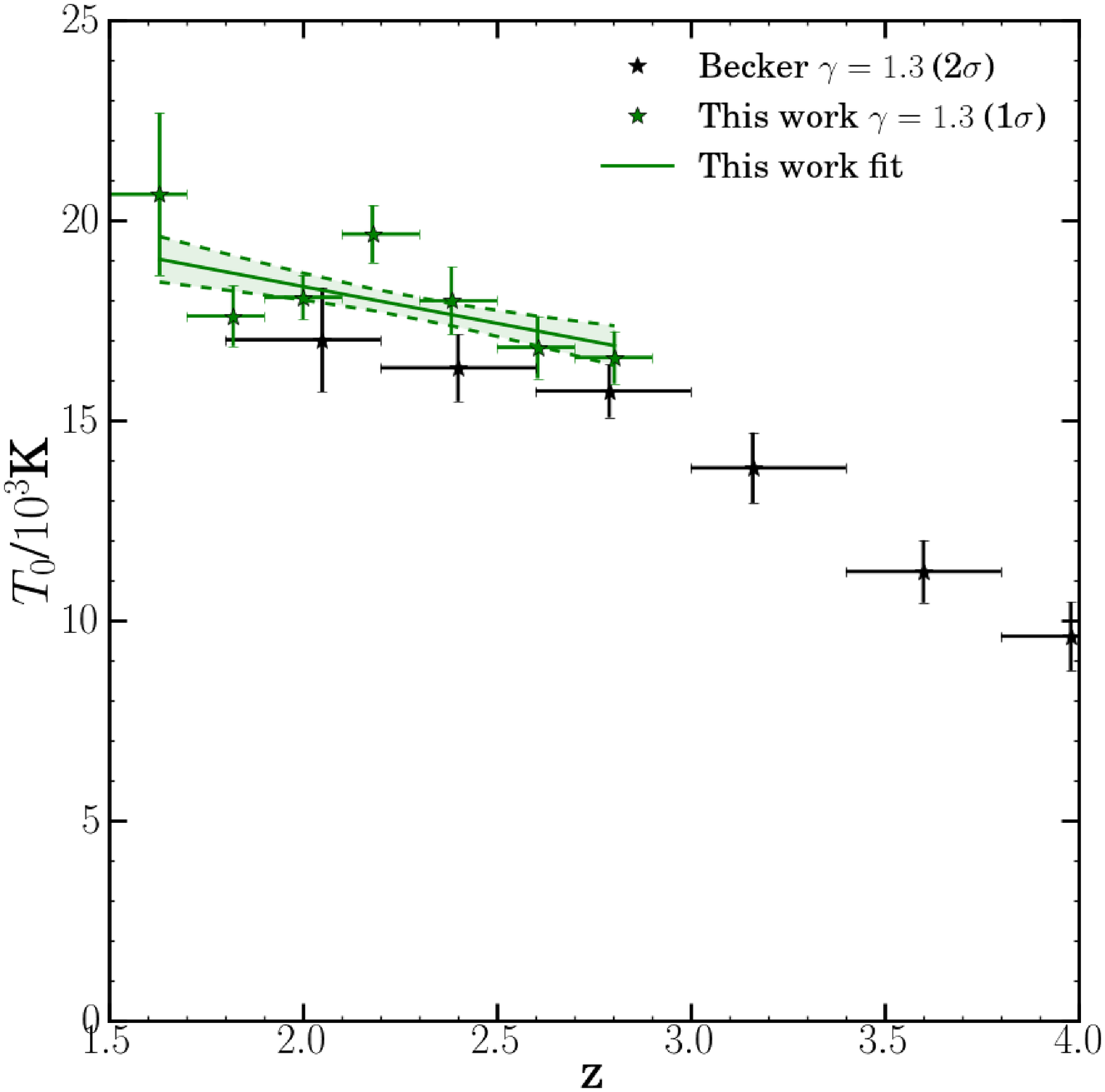}\\

\end{tabular}
\caption{\small Temperature at the mean density, $T_{0}$, inferred from $T(\bar{\Delta})$ in this work (green points/stars) and in the work of \citet{Becker11} (black point/stars) for different assumptions of the parameter $\gamma$: $T_{0}$ for $\gamma\sim 1.5$ (see table 4 for the exact values) (top panel) and for $\gamma=1.3$ (bottom panel).  The linear least square fits of our data points (green line) are presented for both the choices of $\gamma$ with the corresponding $1\sigma$ error on the fit (shaded region). The fits show a change of slope in the temperature evolution: from the increase of $T_{0}$ between $z\sim 4$ and $z\sim 2.8$ there is a tendency of a flattening for $z\lesssim 2.8$ with a decrease in the temperature for $\gamma\sim 1.5$ and a slowdown of the reheating for $\gamma=1.3$. } 

\label{fig:slopes}
\end{figure}

Using the T--$\rho$ relation, characterized by different $\gamma$ values, we translate our measurements of $T(\bar{\Delta})$ to values of temperature at the mean density ($T_{0}$). In Figure \ref{fig:slopes}  we present $T_{0}$ under two different assumptions for $\gamma$: for $\gamma$ values measured from fiducial simulations (A15--G15) in the top panel and for  a constant $\gamma$ =1.3 in the bottom panel. In both the plots our results are compared with those from \citet{Becker11}. 

In the first case we use a $\gamma$ parameter that varies slightly  with redshift, with $\gamma\sim 1.5$ at $z=3$ (the exact values are reported in Table \ref{table:results} and have been recovered from the fiducial simulation C15). These $\gamma$ values are close to the maximum values expected in reality and therefore correspond to the minimum $T_{0}$ case. Our results in the overlapping redshift range ($2.1\lesssim z \lesssim2.9$) are broadly consistent with the previous ones  but show slightly higher values, a difference that can be attributed to the  variation with respect to the previous work in the values of the characteristic overdensities from which we derived our $T_{0}$ values (see Section 6.3). The extension at lower redshifts suggests a tendency of flattening of the increase in $T_{0}$ that can be interpreted as a footprint for the completion of the reheating of the IGM by He II reionization. The least-square linear fit of our data presented in Figure \ref{fig:slopes} (top panel) for this particular choice of $\gamma$ shows, in fact, quantitatively an inversion in the slope of the temperature evolution at the mean density:  the general trend of the temperature is therefore an increase from $z\sim4$ to $z\sim2.8$ with a subsequent flattening of $T_{0}$ around $\sim 12000 $K at $z\sim2.8$. The evolution of the temperature for $z\lesssim2.8$ is  generally consistent with a linear decrease of slope $a =0.80\pm 0.81 (1\sigma)$ generally in agreement with the decrease registered in \cite{Becker11} for the same choice of $\gamma$. 

The situation is similar in the second case for a constant $\gamma=1.3$. This choice, which is motivated by the numerical simulations of \citet{McQuinn09}, corresponds to a mild flattening of the temperature--density relation, as expected during an extended He II reionization process. The trend of $T_{0}$ again shows a strong increase in the temperature from $z\sim 4$ to $z\sim 2.8$ and then a tendency of flattening from $z\sim2.8$ towards lower redshift. However, the temperature obtained in this case is higher, fluctuating around $\sim17000$ K. The scatter between our data points and the ones from \citet{Becker11} is also smaller for this choice of $\gamma$, even if ours are slightly higher on average. In this case the linear fit of our data points at $z\lesssim2.8$ suggests a change in the slope of the temperature evolution but, while in the previous case we register a positive slope, for this $\gamma$ choice we see only a slowdown in the increasing temperature, with a slope that assumes the value $a=-1.84\pm1.06 (1\sigma)$.

The exact redshift of the temperature maximum, reached by the IGM at the mean density approaching the tail-end of He II reionization, is then still dependent on the choice of  $\gamma$, as already pointed out in \cite{Becker11}. Nevertheless, the extension at lower redshift of our data points gives  stronger evidence about the end of this event.  In fact,  while an increase in the temperature for $z\sim4-2.8$  has been recorded in the previous work, if $\gamma$ remains roughly constant, a tendency to a temperature flattening at lower redshift is suggested for both the choice of $\gamma$. A particularly important result is the suggestion of a decrease in $T_{0}$ in the case of  $\gamma\sim1.5$. In fact, according to recent analysis with the line fitting method (\cite{Rudie12}; \cite{Bolton13}) at redshift $z=2.4$ there is good evidence for $\gamma= 1.54\pm0.11(1\sigma)$ and, because we expect that at the end of He II reionization $\gamma$ will tend to come back to the asymptotic value of 1.6, indicating equilibrium between photoionization and cooling due to the adiabatic expansion, the possibility to have $\gamma\lesssim1.5$ for $z\lesssim2.4$ seems to be not realistic. Even if in this work we did not attempt to constrain the temperature--density relation,  the scenario in the top panel of Figure \ref{fig:slopes} seems likely to reproduce the trend in the evolution of the temperature, at least at low redshift, with our results reinforcing the picture of the reheating of the IGM due to He II reionization being almost complete at $z\sim 2.8$, with a subsequent tendency of a cooling, the rate of which will depend on the UV background.

\subsubsection{The UV background at low redshifts} 
The tendency of our $T_{0}$  results to flatten at $z\lesssim2.8$  seems to suggest that at these redshifts the reheating due to the He II reionization has been slowed down, if not completely exhausted, marking the end of this cosmological event.  In the absence of reionization's heating effects the temperature at the mean density of the ionized plasma is expected to approach a thermal asymptote that represents the balance between photoionization heating and cooling due to the adiabatic expansion of the Universe. 
The harder the UV background (UVB) is, the higher the temperature will be because each photoionization event deposits more energy into the IGM. In particular, under the assumption of a power-law ionizing spectrum, $J_{\nu}\propto\nu^{ -\alpha}$, and that He II reionization no longer contributes any significant heating, the thermal asymptote can be generally described by (\citealt{HuiGnedin1997}; \citealt{HuiHaiman03}):
\begin{equation} 
T_{0}= 2.49\times 10^{4} K \times (2+\alpha)^{-{{1}\over{1.7}}} \biggl({{1+z}\over{4.9}}\biggr)^{0.53}  ,
\end{equation}

where the parameter $\alpha$ is the spectral index of the ionizing source. 
The observational value of  $\alpha$ is still uncertain. From direct measurements of QSO rest-frame continua, this value has been found to range between 1.4 and 1.9 depending on the survey (e.g. \citealt{Telfer02}; \citealt{Shull12}) whereas for galaxies the values commonly adopted range between 1 and 3 (e.g. \citealt{Bolton07}; \citealt{Ouchi09}; \citealt{Kuhlen12}) even if, in the case of the emissivity of realistic galaxies, a single power law is likely be consider an oversimplification. 

\begin{figure}
\centering
\includegraphics[width=0.4\textwidth]{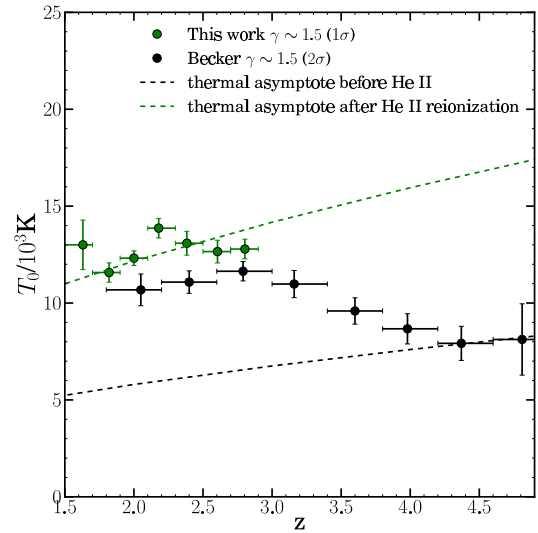}
\caption{\small An example of thermal asymptotes before and after He II reionization with different UVB shapes. The evolution of the thermal asymptote  for the model of  \citet{HuiHaiman03} with $\alpha$ (Eq.~4) scaled to match the high redshift ($z\sim4$--5) data of \citet{Becker11} (black dashed line) is compared with the same model with $\alpha$ scaled to match our results (assuming $\gamma\sim1.5$) at $z\lesssim 2.8$ (green dashed line). The significant change in $\alpha$ required over the redshift range $2.8\lesssim z\lesssim4.5$ in this example suggests that the UV background has changed, hardening during the He II reionization.} 
\label{fig:slopesUV}
\end{figure}

Because our data at $z\lesssim2.8$ do not show any strong evidence for a rapid decrease or increase in the temperature, here we assume that this redshift regime already traces the thermal asymptote in Eq.~4.  Under this hypothesis we can then infer some suggestions about the expectation of a transition of the UV background from being dominated maily by stars to being dominated mainly by quasars over the course of the He II reionization ($2\lesssim z \lesssim5$). In Figure \ref{fig:slopesUV} we show, as an illustrative example only, two models for the thermal asymptote: the first is the model of \citet{HuiHaiman03} for the expected cooling in the absence of He II reionization, with $\alpha$ scaled to 5.65 to match the flattening of the \citet{Becker11} data at $z\sim 4$--5, while in the second case $\alpha$ was scaled to 0.17 to match our results (for $\gamma\sim 1.5$) at $z\lesssim 2.8$. These values are at some variance with the quantitative expectations: in the first case our UVB spectrum is significantly softer compared to typical galaxies-dominated spectrum while, after He II reionization, our value is somewhat harder than a typical quasar-dominated spectrum. We also emphasise that any such estimate of changes in the spectral index also involves the considerable uncertainties, already discussed, connected with the correct position of the peak in $T_{0}$ and the choice of $\gamma$, and so we cannot make firm or quantitative conclusions here. However, in general, the observed cooling at higher temperatures at $z\lesssim 2.8$ seems to suggest that the shape of the UV background has changed, hardening with the increase in temperature during to the reionization event. 


\section{DISCUSSION} 
The main contribution of our work is to add constraints on the thermal history of the IGM down to the lowest optically-accessible redshift, $z\sim1.5$. These are the first temperature measurements in this previously unexplored redshift range. In this section we discuss the possible implications of our results in terms of compatibility with theoretical models.

Measuring the low-redshift thermal history is important for confirming or ruling out the photo-heating model of HeII reionization and the new blazar heating models. According to many models of the former, the HeII reionization should have left a footprint in the thermal history of the IGM: during this event, considerable additional heat  is expected to increase the temperature at the mean density of the cosmic gas ($T_{0}$) at $z\lesssim 4$ \citep{HuiGnedin1997}.  The end of He II reionization is then characterized by a cooling  of the IGM due to the adiabatic expansion of the Universe with specifics that will depend on the characteristics of the UV background.  However, even if some evidence has been found for an increase in the temperature at the mean density from $z\sim4$ down to $z\sim2.1$ (e.g. \citealt{Becker11}), the subsequent change in the evolution of $T_{0}$ expected after the end of the He II reionization has not been clearly characterized yet and remains strongly degenerate with the imprecisely constrained slope of the temperature--density relation $\gamma$ (see Eq. 1). This, combined with several results from PDF analysis which show possible evidence for an inverted temperature--density relation (\citealt{Becker07}; \citealt{Bolton08}; \citealt{Viel09}; \citealt{Calura12}; \citealt{Garzilli12}), brought the development of a new idea of volumetric heating from blazar TeV emission (\citealt{Chang12}; \citealt{Puchwein12}). These models, where the heating rate is independent of the density, seem to naturally explain an inverted T--$\rho$ relation at low redshift. Predicted to dominate the photo-heating for $z\lesssim3$, these processes would obscure the change in the  temperature evolution trend due to the HeII reionization, preventing any constraint on this event from the thermal history measurements.  

A main motivation for constraining the temperature at lower redshifts than $z\sim2.1$ is to confirm evidence for a flattening in the already detected trend of increasing temperature for $z\lesssim 4$. A precise measurement of a change in the $T_{0}(z)$ slope, in fact, could bring important information about the physics of the IGM at these redshifts and the end of the He II reionization event. It is therefore interesting that our new temperature measurements in Figure \ref{fig:slopes}, which extend down to redshifts $z\sim1.5$, show some evidence for such a change in the evolution for $z \lesssim 2.8$. However, in order to make a fair comparison with different heating models in terms of the temperature at the mean density, we must recognize the fact that we do not have yet strong constraints on the evolution of the T--$\rho$ relation slope as a function of redshift: assuming a particular choice of $\gamma$ for the translation of the temperature values at the characteristic overdensities to those at the mean density, without considering the uncertainties in the slope itself, could result in an unfair comparison. Furthermore, the blazar heating models' T--$\rho$ relation at each redshift can be parametrized with a power-law (of the form of Eq.1) only for a certain range of overdensities that may not always cover the range in our characteristic overdensities (\citealt{Chang12}; \citealt{Puchwein12}).

Therefore, to compare our results with the blazar heating model predictions, we decided to use directly the $T(\bar{\Delta})$ values probed by the forest.

\begin{figure} 
\centering
\includegraphics[width=0.5\textwidth]{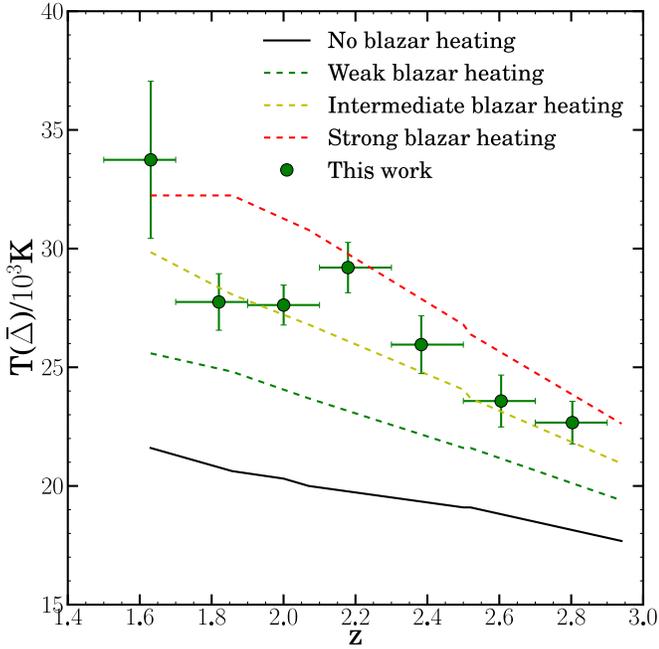} 
\caption{\small Comparison of blazar heating models: the temperature values at the redshift-dependent characteristic overdensities, $T(\bar{\Delta})$, inferred in this work (green points) are compared with the model without a blazar heating contribution (black solid line) and the weak (green dashed line), intermediate (yellow dashed line) and strong (red dashed line) blazar heating models of  \citet{Puchwein12}. The blazar heating predictions were computed at the corresponding using the $\bar{\Delta}(z)$ in Table 4 to order to allow a fair comparison with our $T(\bar{\Delta})$ measurements. Our observational results seem to be in reasonable agreement with the intermediate blazar heating model. The vertical error bars represent the 1-$\sigma$ errors on the temperature measurements.} 

\label{fig:BB2}
\end{figure}

In Figure \ref{fig:BB2} we compare the model without blazar heating contributions, and the weak, intermediate and strong blazar heating models of \citet{Puchwein12}, with our new results for the temperature at the characteristic overdensities. The temperature values for all the models were obtained by computing the maximum of the temperature distribution function at the corresponding redshift-dependent characteristic overdensities [$\bar{\Delta}$] in Table 4 by E.~Puchwein (private communication). Each of the three blazar heating models has been developed using different heating rates, based on observations of 141 potential TeV blazars, under the assumption that their locally-observed distribution is representative of the average blazars distribution in the Universe. The variation in the heating rates is due to the tuning of a coefficient in the model that corrects for systematic uncertainties in the observations: the lower this multiplicative coefficient, the weaker the heating rate (\citealt{Chang12}; \citealt{Puchwein12}). The observational constraints on the IGM thermal evolution in Fig.~\ref{fig:BB2} seem to be in reasonable agreement with the intermediate blazar heating model, even if some fluctuations toward higher temperatures reach the range of values of the strong blazar model. This result in general reflects what was found by \citealt{Puchwein12} (their figure 5) in their comparison with the temperature measurements of \citet{Becker11}, even if in that case the models were tuned to different $\bar{\Delta}$ values than ours. This general agreement could be explained by the fact that our $T(\bar{\Delta})$ measurements closely match those of \citet{Becker11} in the common redshift range $2.0\lesssim z\lesssim 2.6$ (see Fig.~\ref{fig:TD}) and by the weak dependency on the density of the blazar heating mechanism.

According to Fig.~\ref{fig:BB2}, the model from \citet{Puchwein12} without a blazar heating contribution, that is based on the UV background evolution of \citet{Faucher09}, shows temperature values significantly lower than our ($\bar{\Delta})$ measurements $z\lesssim 3$. However, this model does not take into account the contribution of the diffuse hard X-ray background (\citealt{Churazov07}). The excess energy of these ionizing photons could, in fact, contribute to the heating, shifting the range of temperatures towards higher values. Also, to definitely ruled out or confirm any of the different thermal histories, an interesting further test would be to compare the temperature at the mean density between observations and models. 

That is, there is a strong need for model-independent measurements of $T_{0}$ that would allow a straightforward comparison between different T--$\rho$ relations, and so different model predictions. In this context a promising prospect is the use of the He II Lyman-$\alpha$ forest for the identification of the absorption features useful for an improved line-fitting constraint of $\gamma$. The possibility to calculate directly the ratio between $b$ parameters of the corresponding H I and He II lines would, in fact, make possible an easier and more precise selection of the lines that are dominated by thermal broadening ($b_{H I}/b_{He II}\simeq2$). Even if the low S/N of the UV spectra currently available make this identification difficult (\citealt{Zheng04}), possible future, space-based telescopes UV with high resolution spectrographs 
(e.g. \citealt{Postman09}) may offer the opportunity to improve the quality and the number of ``clean"  He II forests for future analyses.

\section{CONCLUSIONS}

 In this work we have utilized a sample of 60 VLT/UVES quasar spectra to make a new measurement of the IGM temperature evolution at low redshift, $1.5\la z\la2.8$, with the curvature method applied to the H{\sc \,i} Lyman-$\alpha$ forest. For the first time we have pushed the measurements to the lowest optically-accessible redshifts, $z\sim1.5$.
Our new measurements of the temperature at the characteristic overdensities traced by the Ly-$\alpha$ forest, $T(\bar{\Delta})$, are consistent with the previous results of Becker et al.~(2011) in the overlapping redshift range, $2.0 < z < 2.6$, despite the datasets being completely independent. They show the same increasing trend for $T(\bar{\Delta})$ towards lower redshifts while, in the newly-probed redshift interval $1.5\la z\la2.0$, the evolution of $T(\bar{\Delta})$ is broadly consistent with the extrapolated trend at higher redshifts.

The translation of the $T(\bar{\Delta})$ measurements into values of  temperature at the mean density, $T_{0}$, depends on the slope of the temperature--density relation, $\gamma$, which we do not constrain in this work. However, for reasonable, roughly constant, assumptions of this parameter, we do observe some evidence for a change in the slope of the temperature evolution for redshifts $z\lesssim2.8$, with indications of at least a flattening, and possibly a reversal, of the increasing temperature towards lower redshifts seen in our results and those of \citet{Becker11} for $2.8\lesssim z\lesssim 4$. In particular, for the minimum $T_{0}$ case, with $\gamma\sim 1.5$, the extension towards lower redshifts provided by this work adds to existing evidence for a decrease in the IGM temperature from $z\sim2.8$ down to the lowest redshifts probed here, $z\sim1.5$. This could be interpreted as the footprint of the completion of the reheating process connected with the He II reionization.

Following the additional hypothesis that our low redshift temperature measurements are already tracing the thermal asymptote, the cooling of $T_{0}$ inferred at $z\lesssim2.8$ (assuming $\gamma\sim 1.5$) may suggest that the UV background has changed, hardening during the He II reionization epoch. However, the expectation for the evolution of $T_{0}$ following HeII reionization will depend on the evolution in $\gamma$ and on details of the reionization model.

We also compared our T($\bar{\Delta}$) measurements with the expectations for the models of \citet{Puchwein12} with and without blazar heating contributions. To allow a fair comparison with our observed values, the model predictions were computed at the corresponding (redshift-dependent) characteristic overdensities ($\bar{\Delta}$). Our observational results seem to be in reasonable agreement with a moderate blazar heating scenario. However, to definitely confirm or rule out any specific thermal history it is necessary to obtain new, model-independent measurements of the temperature at the mean density.

With the IGM curvature now constrained from $z\sim4.8$ down to $z\sim1.5$, the main observational priority now is clearly to tightly constrain the slope of the temperature--density relation, $\gamma$, and its evolution over the redshift range $1.5\la z\la 4$. This is vital in order to fix the absolute values of the temperature at the mean density and to comprehensively rule out or confirm any particular heating scenarios.

Finally, we note that, even though our new measurements have extended down to $z\sim1.5$, there is still a dearth of quasar spectra with high enough S/N in the 3000--3300\,\AA\ spectral range to provide curvature information in our lowest redshift bin, $1.5<z<1.7$. We have searched the archives of both the VLT/UVES and Keck/HIRES instruments for new spectra to contribute to this bin. However, the few additional spectra that we identified had relatively low S/N and, when included in our analysis, contributed negligibly to the final temperature constraints. Therefore, new observations of UV-bright quasars with emission redshifts $1.5\la z_{\rm em}\la1.9$ are required to improve the temperature constraint at $1.5<z<1.7$ to a similar precision as those we have presented at $z>1.7$.

\section*{Acknowledgements}

We thank E.~Puchwein for providing us with the temperature at our characteristic overdensity values within the various blazar heating scenarios. EB thanks E.~Tescari for several useful discussions. We thank the referee, M. Shull, for helpful comments that clarified several points in the paper. The hydrodynamical simulations used in this work were performed using the Darwin Supercomputer of the University of Cambridge High Performance Computing Service (http://www.hpc.cam.ac.uk/), provided by Dell Inc.~using Strategic Research Infrastructure Funding from the Higher Education Funding Council for England.  MTM thanks the Australian Research Council for \textit{Discovery Project} grant DP130100568 which supported this work. GDB gratefully acknowledges support from the Kavli Foundation. JSB acknowledges the support of a Royal Society University Research Fellowship.


\appendix

\section{The effect of the optical depth calibration on the temperature measurements}

In this section we demonstrate that, in the curvature analysis of a particular sample of sight-lines, calibrating the Lyman-$\alpha$ forest simulations with the effective optical depth of the sample provides robust measurements of the temperature at the characteristic overdensities. We will show how, even if different $\tau_{\rm eff}$ calibrations produce different characteristic overdensities [$\bar{\Delta}(z)$; see Section 6.3], the temperature measurements, $T(\bar{\Delta})$, at each redshift will not be affected significantly by systematic effects related to possible biases in the sample selection. Instead, discrepancies in the characteristic overdensities will shift the derived temperature at the mean density. Nevertheless, this effect in the $T_{0}$ values will be modest, causing a disparity at the level of the observational 1-$\sigma$ error bars.

The test can be summarized as follows. We randomly select two sub-samples of 300 spectral sections from the suite of simulations of one thermal history. One of the sub-samples is selected in a biased way to result in a higher effective optical depth than the other; this difference is designed to be similar to that observed between the UVES sample used in this work and the sample adopted in the previous work of \citet{Becker11}. Treating the two sub-samples as observational data, we analyse them separately with the curvature method presented in Section 4 and obtain the corresponding $T(\bar{\Delta})$ and $T_{0}$ measurements. The two sets of $T(\bar{\Delta})$ are found not to differ significantly, while a modest shift in the $T_{0}$ values is observed due to discrepancies in the recovered $\bar{\Delta}$ values at each redshift. The details of this test are described below.

\subsection{Selection of synthetic sub-samples}

\begin{figure}
\centering
\includegraphics[width=0.5\textwidth]{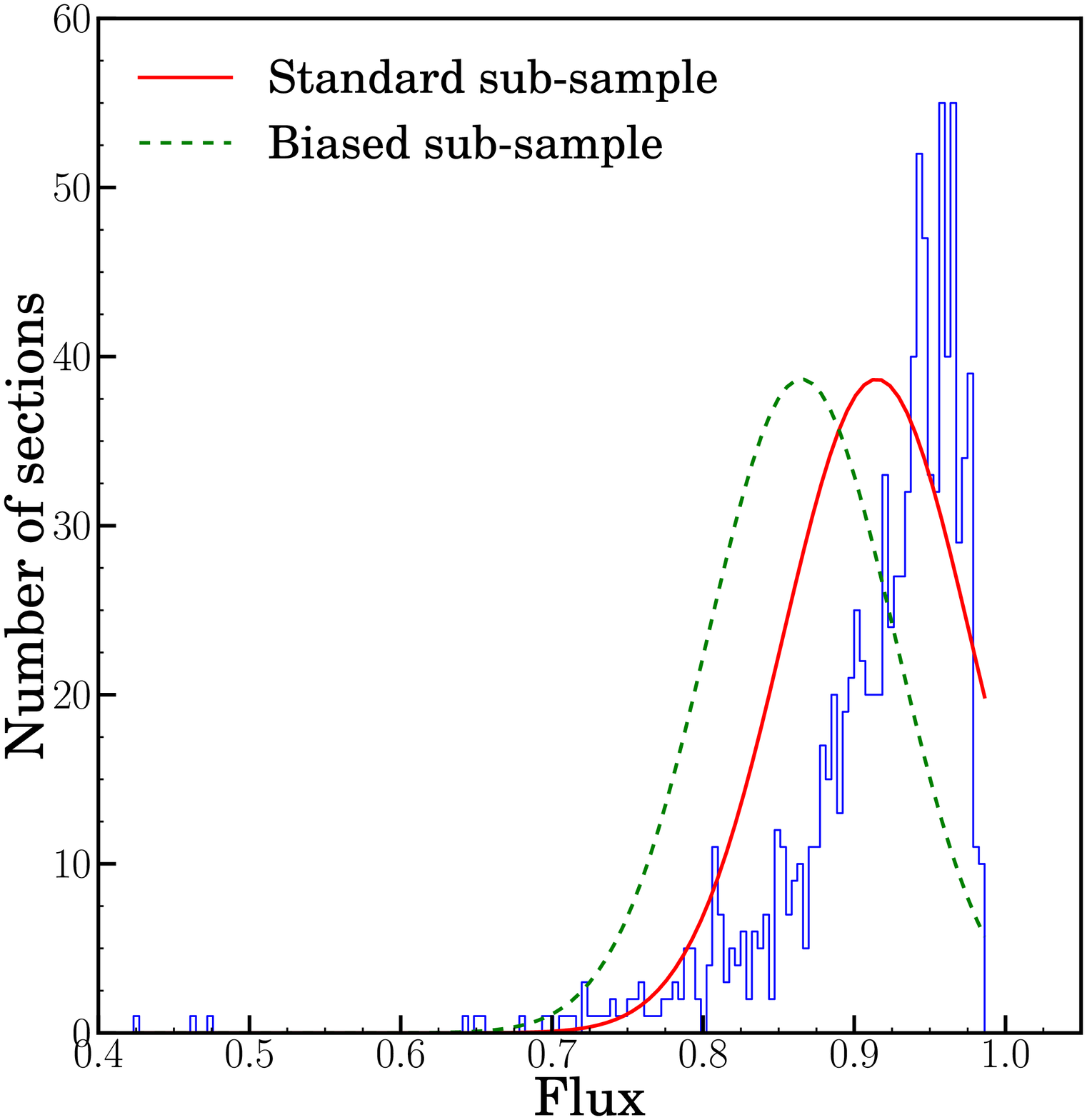} 
\caption{\small Global distribution of the mean fluxes at $z=1.75$ for all the 1024 sections of the simulation C15. Two sub-samples have been selected following the probability distributions of the two Gaussian curves. The solid red curve was fit directly to the flux distribution and was used to select the standard sub-sample. The dashed green curve has the same FWHM but its mean value was shifted toward lower fluxes to allow the selection of the biased sub-sample with a higher mean optical depth.} 
\label{fig:Selection1}
\end{figure}

We chose the 1024 synthetic sections of our fiducial simulation C15 (see Table 4) as the ``global'' sample from which to select, at each redshift, two sub-samples of $\sim 300$ sections with two slightly different mean $\tau_{\rm eff}$ that would simulate two random observational samples. We deliberately biased the mean optical depth of the second sub-sample towards higher values using the method explained in Fig.~\ref{fig:Selection1}: we fit a Gaussian function to the global distribution of mean fluxes (at $z=1.75$ in the example shown in the Figure) from all 1024 sections of the C15 simulation and used this, and a shifted version of it, as the probability distributions for selecting sections randomly for the two sub-samples of 300 sections each. By construction, the first subsample -- which we call the `standard sub-sample' for clarity -- will have a mean $\tau_{\rm eff}$ very close to the global mean. However, the mean optical depth of the second sub-sample -- called the `biased sub-sample' -- is selected from the same probability distribution shifted slightly to lower mean fluxes, so it results in a higher mean $\tau_{\rm eff}$. The shift in the probability distribution was tuned so that difference in the mean $\tau_{\rm eff}$ at each redshift reflected the difference observed between our real UVES sample and the data used by \citet{Becker11}.

\begin{figure} 
\centering
\includegraphics[width=0.5\textwidth]{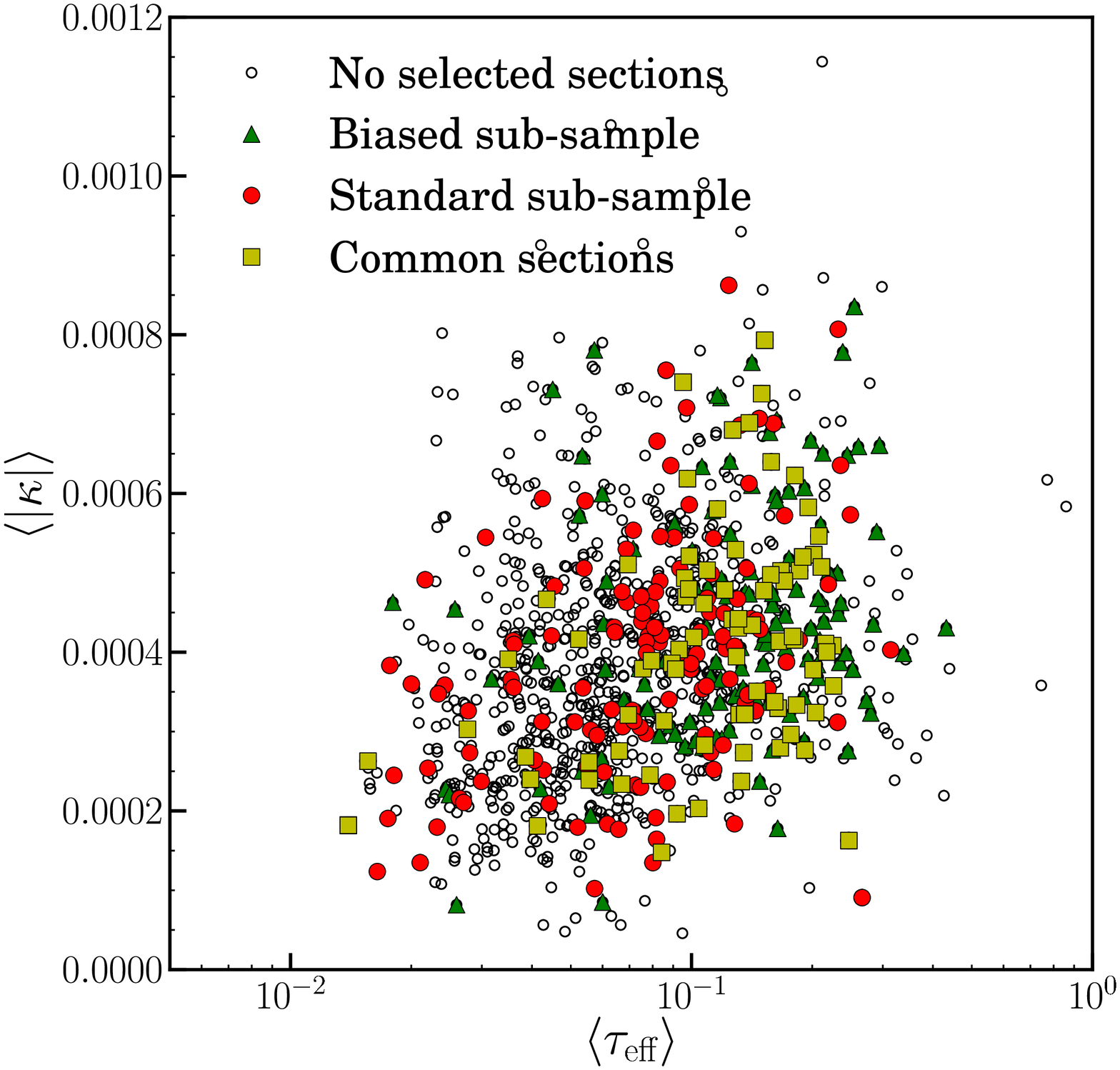} 
\caption{\small Example of the $\langle|\kappa|\rangle$ distribution as a function of the effective optical depth of the spectral sections at $z=1.75$ of the simulation C15. The open black circles represent synthetic sections and the different colored solid symbols show the sub-sample selection results: red points represent the standard sub-sample while the green triangles represent the biased sub-sample with higher mean optical depth. Sections that fall in both sub-samples are also indicated as yellow squares.}
\label{fig:Selection2}
\end{figure}

The results of the sub-sample selection are presented in Fig.~\ref{fig:Selection2} where we show the positions on the absolute curvature--mean $\tau_{\rm eff}$ plane of all the sections from simulation C15 at $z=1.75$. As expected, the distribution of mean $\tau_{\rm eff}$ and curvature in the standard sub-sample is very similar to the parent distribution. Also, while noting that, by construction, the biased sub-sample has a higher mean $\tau_{\rm eff}$ than the standard sub-sample, we also see that the mean curvature of the biased sub-sample is very similar to the parent distribution. The sub-sample selection therefore should allow a test of the effect of selecting an observational sample with a higher mean $\tau_{\rm eff}$ on the measured $T(\bar{\Delta})$ values.

\subsection{Parallel curvature analysis and results}

After the selection of the two sub-samples we treated them as two separate observational datasets and we analysed the curvature following the steps presented in Section 4. That is, at each redshift and for each sub-sample, we computed the $\langle|\kappa|\rangle$ values and measured $T(\bar{\Delta})$ after calibrating all the simulations (from all thermal histories) with the mean $\tau_{\rm eff}$ found in that particular sub-sample. Finally, using the T--$\rho$ relation, we derived the values of $T_{0}$ under the assumption of $\gamma=1.54$  (corresponding to the chosen thermal history C15). 

\begin{figure} 
\centering
\includegraphics[width=0.5\textwidth]{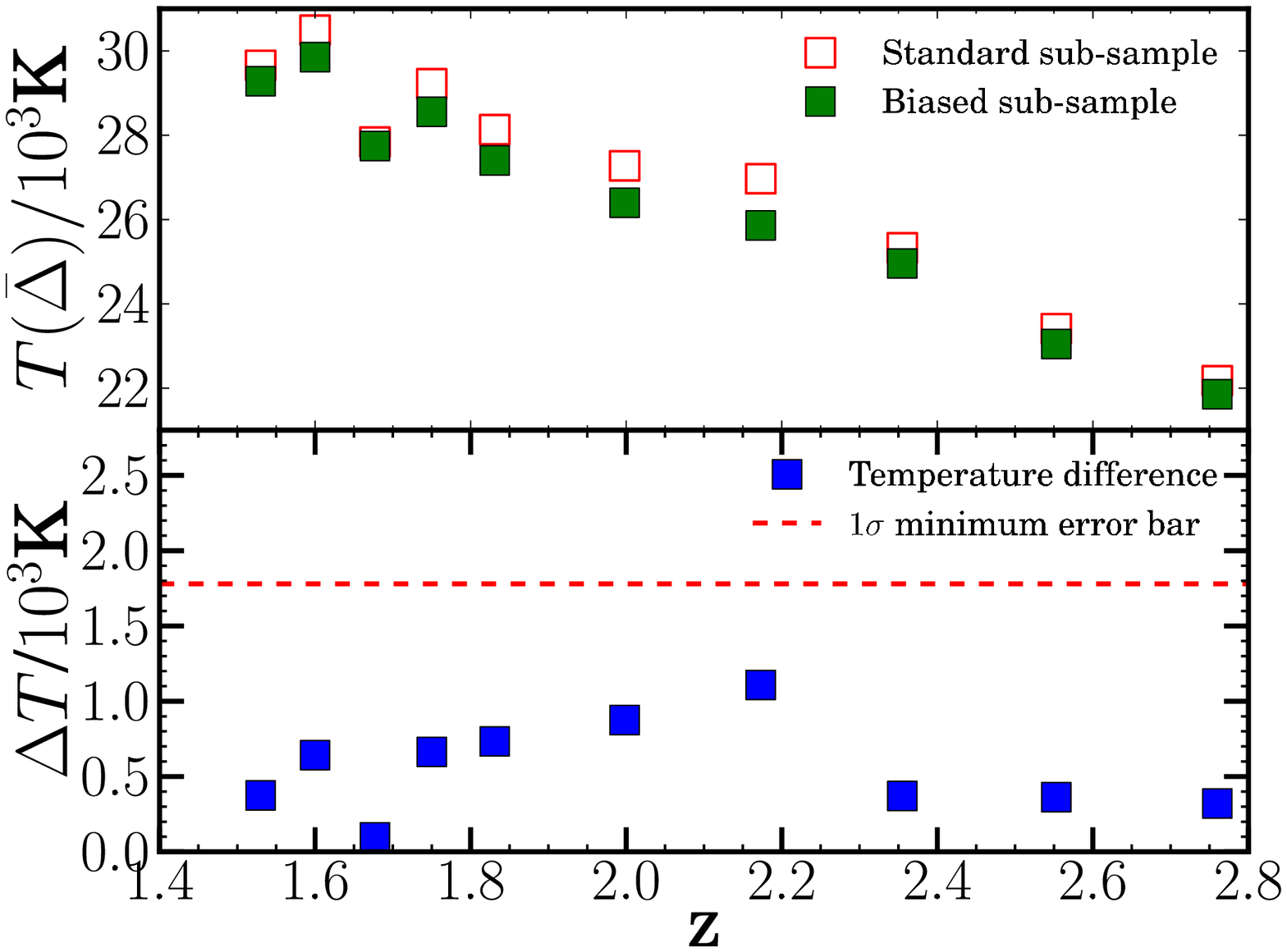} 
\caption{\small Uppler panel: $T(\bar{\Delta})$ values computed from the curvature analysis of the two synthetic sub-samples. Lower panel: Difference in $T(\bar{\Delta})$ between the standard and biased sub-samples, $\Delta T$. The discrepancy between the temperature values is shown as a function of redshift (blue squares) and the minimum 1$\sigma$ error bar observed in the UVES sample in this work (see Table 4) is given by the red dashed line for comparison.} 
\label{fig:Tdifference}
\end{figure}

As expected, calibrating the simulations with the two different effective optical depths gave slightly different values for the $\bar{\Delta}$ at each redshift of the two sub-samples. However, we find excellent agreement between the $T(\bar{\Delta})$ values for the two sub-samples, as shown in the top panel of Fig.~\ref{fig:Tdifference}. There we also plot the difference between the $T(\bar{\Delta})$ values, $\Delta T$, from the two sub-samples at each redshift in the lower panel. This difference is $\Delta T\lesssim 1100$\,K at all redshifts and typically than 800\,K. Considering that the temperature measurements at the characteristic overdensities presented in this work have a minimum $1\sigma$ error bar of $\sim 1800$,K, these $\Delta T$ all fall inside the current statistical uncertainty budget. Also, we expect small, non-zero values of $\Delta T$, and small variations with redshift, due to the sample variance connected with the selection of the sub-samples. In general, we can conclude that calibration of the simulations with the effective optical depth of the particular observational sample being analysed results in a self-consistent measurement of $T(\bar{\Delta})$.

\begin{figure} 
\centering
\includegraphics[width=0.5\textwidth]{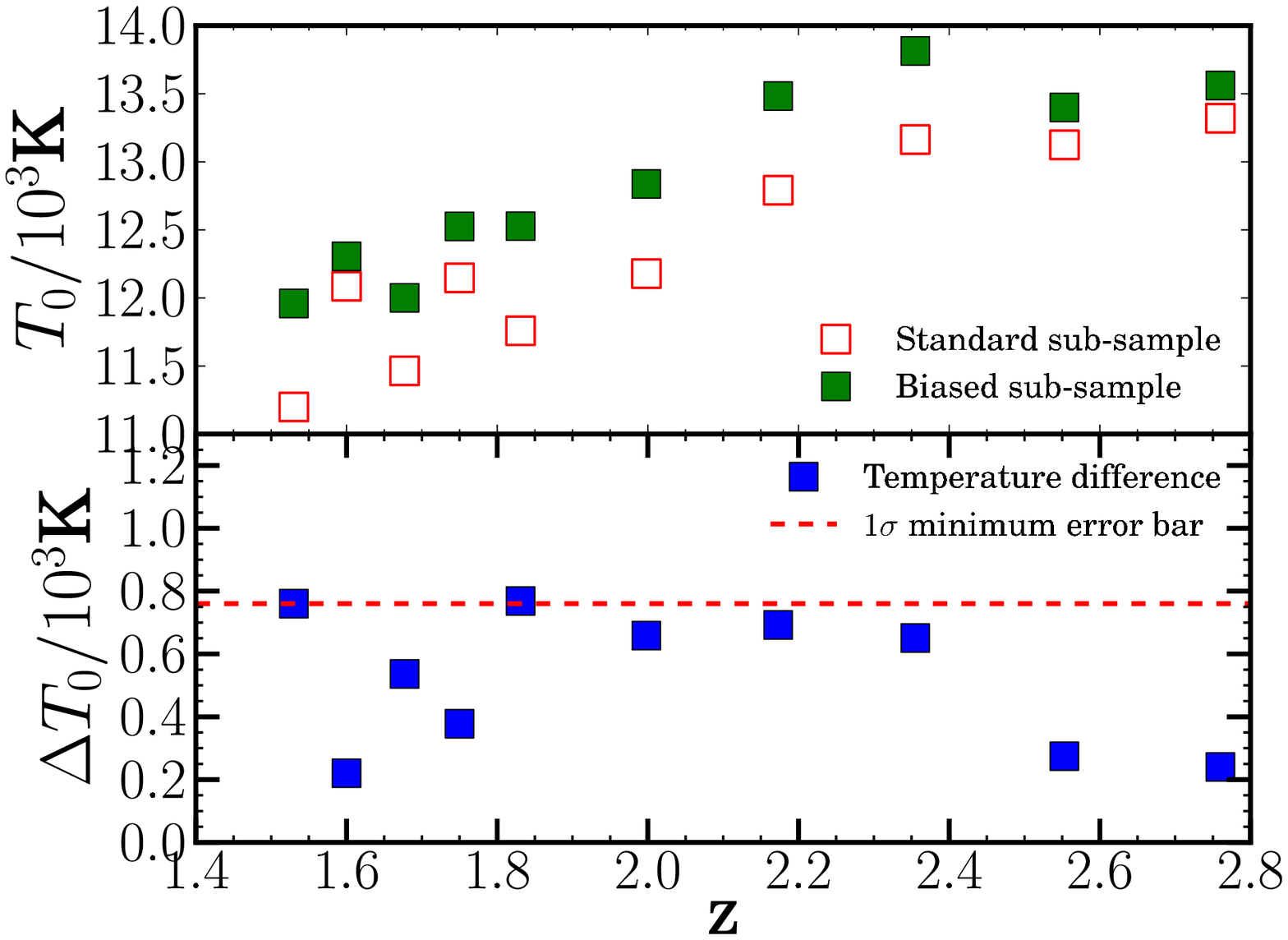} 
\caption{\small Uppler panel: $T_{0}$ values computed from the $T(\bar{\Delta})$ measurements of the two synthetic sub-samples under the assumption of $\gamma=1.54$. Lower panel: Difference in $T_{0}$ between the standard and biased sub-samples, $\Delta T_{0}$. The discrepancy between the temperature values is shown as a function of redshift (blue squares) and the minimum 1-$\sigma$ error bar observed in the UVES sample in this work (see Table 4) is given by the red dashed line for comparison.} 
\label{fig:T0difference}
\end{figure}

In terms of the temperature values at the mean density, we find that the biased sub-sample produces a systematically higher temperature, as expected (top panel of Fig.~\ref{fig:T0difference}). This discrepancy is due to the slightly different values of $\bar{\Delta}$ at each redshift in the two different sub-samples. However, it is still modest and is generally below the minimum 1-$\sigma$ uncertainty of the measurements presented in this work (lower panel in Fig.~\ref{fig:T0difference}).

\label{lastpage}

\end{document}